\documentclass[aps,reprint,superscriptaddress,amsmath,amssymb,twocolumn]{revtex4-1}

\usepackage{graphicx} 
\usepackage{dcolumn} 
\usepackage{bm} 
\usepackage{amsmath}
\usepackage{braket}
\usepackage{natbib}
\usepackage{color}%
\usepackage{ulem}
\usepackage{lineno}

\newcommand*\patchAmsMathEnvironmentForLineno[1]{%
  \expandafter\let\csname old#1\expandafter\endcsname\csname #1\endcsname
  \expandafter\let\csname oldend#1\expandafter\endcsname\csname end#1\endcsname
  \renewenvironment{#1}%
     {\linenomath\csname old#1\endcsname}%
     {\csname oldend#1\endcsname\endlinenomath}}%
\newcommand*\patchBothAmsMathEnvironmentsForLineno[1]{%
  \patchAmsMathEnvironmentForLineno{#1}%
  \patchAmsMathEnvironmentForLineno{#1*}}%
\AtBeginDocument{%
\patchBothAmsMathEnvironmentsForLineno{equation}%
\patchBothAmsMathEnvironmentsForLineno{align}%
\patchBothAmsMathEnvironmentsForLineno{flalign}%
\patchBothAmsMathEnvironmentsForLineno{alignat}%
\patchBothAmsMathEnvironmentsForLineno{gather}%
\patchBothAmsMathEnvironmentsForLineno{multline}%
}

\begin{document}


\title{Conversion between electron spin and microscopic atomic rotation}




\author{Masato Hamada}
\affiliation{Department of Physics, Tokyo Institute of Technology, 2-12-1 Ookayama, Meguro-ku, Tokyo 152-8551, Japan}
\author{Shuichi Murakami}
\affiliation{Department of Physics, Tokyo Institute of Technology, 2-12-1 Ookayama, Meguro-ku, Tokyo 152-8551, Japan}
\affiliation{TIES, Tokyo Institute of Technology, Ookayama, Meguro-ku, Tokyo 152-8551, Japan}



\date{\today}

\begin{abstract}
We theoretically investigate the microscopic mechanism of  conversion between the electron spin and the microscopic local rotation of atoms in crystals. 
In phonon modes with angular momenta, the atoms microscopically rotate around their equilibrium positions in crystals. 
In a simple toy model with phonons, we calculate the spin expectation value by using the adiabatic series expansion.
We show that the time-averaged spin magnetization is generated by the microscopic local rotation of atoms via the spin-orbit interaction. On the other hand, in the system with a simple vibration of atoms, time-averaged spin magnetization becomes zero due to the time-reversal symmetry.
Moreover, the magnitude of the time-averaged spin magnetization depends on the inverse of the difference of instantaneous eigenenergy, and we show that it becomes smaller in band insulators with a larger gap.
\end{abstract}



\maketitle



\section{Introduction}

The conversion between the magnetization and the mechanical rotation, such as Einstein-de Haas effect~\cite{Einstein} and Barnett effect~\cite{Barnett}, has been known since more than a hundred years ago, and this has been focused on in spintronics recently.
These effects have been understood by means of the spin-rotation coupling~\cite{SRC01, SRC02, SRC03}, which means that the mechanical rotation acts on electron spins as an effective magnetic field.
In spintronics, mechanical generation of spin current has been theoretically proposed or experimentally reported by several means, such as the surface acoustic wave~\cite{SRC01,PhysRevLett.119.077202}, the twisting vibrational mode in carbon nanotubes~\cite{Masato01}, and the flow of the liquid metals~\cite{SHD01}, via the spin-rotation coupling.
However, microscopic mechanism of conversion between electron spins and the rotational motion of atoms in solids is not well known.

Recently, an angular momentum of phonons, which is a microscopic local rotation of atoms in crystals, has been formulated~\cite{PAM01}, and various phenomena related with it have been energetically investigated, such as the phonon Hall effect~\cite{PHEexp01, PHEexp02, PHEexp03,PhysRevLett.100.145902,PhysRevLett.105.225901}, spin relaxation~\cite{PAM02,PhysRevB.97.174403,PhysRevLett.121.027202}, the orbital magnetic moment of phonons~\cite{PhysRevMaterials.3.064405}, and conversion between phonons and magnons~\cite{NatPhys.14.500,PhysRevB.92.214437}. 
In a system without inversion symmetry, the phonons have chirality~\cite{PAM03}. 
In particular, at the valleys in momentum space, the angular momentum of phonons is quantized, and such phonons are called chiral phonons. 
Chiral phonons can be excited by a circularly polarized light via intervalley scattering of electrons, and this has been experimentally observed in monolayer tungsten diselenide~\cite{chiralphononexp01}.
Methods of generation of the phonon angular momentum have been proposed, such as a magnetic field~\cite{PAM01, PhysRevMaterials.1.014401}, a Coriolis force in rotating frame~\cite{Wang_2015, PhysRevB.96.064106}, a circularly polarized light~\cite{PAM03, chiralphononexp01}, an infrared excitation~\cite{PhysRevMaterials.3.064405, nova2017}, a temperature gradient in the system without inversion symmetry~\cite{Masato02}, and an electric field in the magnetic insulator~\cite{Masato03}.
Here, one question arises whether electron spins couple with a microscopic local rotation of atoms in crystals.
Because the microscopic local rotation is quite different from the macroscopic rotation studied in the Einstein-de Haas effect and the Barnett effect, it is a crucial issue whether one can apply the spin-rotation coupling to a coupling between the electron spins and a microscopic local rotations.

In this paper, we theoretically propose microscopic mechanism of conversion between an electron spin and a microscopic local rotation of atoms in crystals.
We first introduce a simple toy model, which is a two-dimensional system forming the honeycomb lattice with a microscopic local rotation of atoms.
We set the rotation of atoms to be periodic in time, and we calculate the spin expectation value.
Here we assume that the rotational motion is much slower than the characteristic energy scale of electrons, and we use the adiabatic series expansion in two previous theories; Berry phase in adiabatic process~\cite{GOM} and the iteration of the unitary transformation~\cite{irbm01,APT01} 
in order to calculate a time-averaged spin expectation value.
By using the adiabatic series expansion, we show that the microscopic local rotation of atoms leads to a nonzero spin magnetization even after averaging over one period in our toy model. 
Meanwhile a simple vibration of atoms induces spin magnetization as a function of time, but after time averaging it becomes zero.
Moreover, we discuss dependence on the parameters for time-averaged spin magnetization.

\section{Toy model with phonon angular momentum}
\begin{figure}
\includegraphics[width=8.5cm]{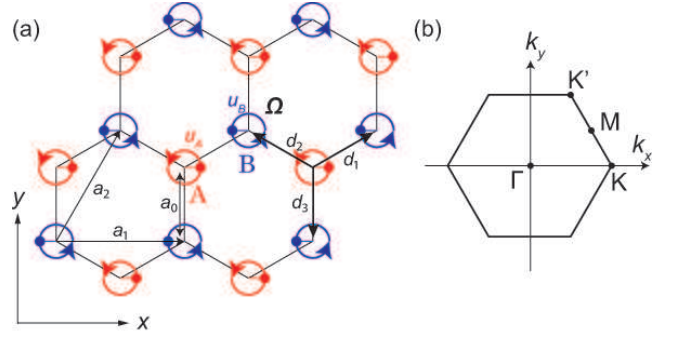}
\caption{\label{dynhoney} 
Schematic figures of  (a) the honeycomb lattice with a microscopic local rotation and (b) the first Brillouin zone.
}
\end{figure}

We introduce a two-dimensional tight-binding model on the honeycomb lattice with a microscopic local rotation of atoms as shown in Fig.~\ref{dynhoney}(a). 
In Fig.~\ref{dynhoney}(b), we show the first Brillouin zone.
The two-dimensional honeycomb lattice without inversion symmetry has chiral phonons at the valleys $K$ and $K^\prime$ in the momentum space~\cite{PAM03}. 
The $K$ and $K^\prime$ phonons have the chirality $\pm 1$, and in these modes, the atoms circularly rotate.
On the other hand, at the zone center $\Gamma$, the longitudinal and the transverse optical phonon modes are degenerate and these modes do not have phonon angular momentum. 
However, we can get the circular polarized phonon modes by superposition of these degenerate modes at the $\Gamma$ point.
Here, we focus on a microscopic local rotation of atoms corresponding to the $\Gamma$-phonon by superposition of two optical degenerate modes.
In Fig.~\ref{dynhoney}(a), the red and blue balls represent the A and B atoms at the two sublattices respectively, and the atoms A and B circularly rotate in the $xy$-plane, with their phases of rotations different by $\pi$.
The primitive vectors are written as $\bm{a}_1 = a(1,0)$, $\bm{a}_2 = a(1/2,\sqrt{3}/2)$ with a lattice constant $a$,
 and the vectors between the nearest neighbor atoms are written as $\bm{d}_1 = a_0(\sqrt{3}/2,1/2)$, $\bm{d}_2=a_0(-\sqrt{3}/2,1/2)$, $\bm{d}_3=a_0(0,-1)$ with the nearest neighbor bond length $a_0 = a/\sqrt{3}$. 
We set the angular velocity of the atoms A and B to be $\Omega$, and the displacement vectors of atoms A and B at time $t$ are represented as
\begin{align}
\bm{u}_{\rm A} = u_0^{\rm A}(\cos \Omega t, \sin \Omega t),~\bm{u}_{\rm B} = -u_0^{\rm B} (\cos \Omega t, \sin \Omega t),
\end{align}
and the displacement vector from the atom A to the atom B is
\begin{equation}
\bm{u} = \bm{u}_{\rm B} - \bm{u}_{\rm A} = -u_{+}(\cos \Omega t, \sin \Omega t)
\end{equation}
with $u_{+} = u_0^{\rm A} + u_{0}^{\rm B}$.

Next, we consider the following nearest-neighbor tight-binding Hamiltonian for electrons, which is similar to the Kane-Mele model , representing a two-dimensional topological insulator~\cite{Z2},
\begin{equation}
\hat{H}_0 = t_0\sum_{\braket{ij}}\hat{c}_{i}^{\dagger}\hat{c}_{j} + \lambda_{\rm v} \sum_{i}\xi_i \hat{c}_{i}^{\dagger}\hat{c}_{i} + i\frac{\lambda_{\rm R}}{a_0} \sum_{\braket{ij}} \hat{c}_{i}^{\dagger}(\bm{s}\times \bm{d}_{ij})_z \hat{c}_j. 
\end{equation}
The first term is a nearest-neighbor hopping term and $t_0$ is a hopping parameter. 
The second term is a staggered potential, which we include to break inversion symmetry. $\xi_{i}$ is $\xi_{\rm A(B)} = +1(-1)$ for the A and B atoms and $\lambda_{\rm v}$ is an on-site potential. 
The third term is a nearest-neighbor Rashba term, which leads to spin-split band structure, and $\lambda_{\rm R}$ is Rashba parameter, $\bm{d}_{ij}$ is the nearest neighbor bond vector, and $s_{\alpha}$ is the Pauli matrices of electron spin.
The operators $\hat{c}_i^\dagger = (c_{i\uparrow}^{\dagger},\hat{c}_{i\downarrow}^{\dagger})$, $\hat{c}_{i} = (\hat{c}_{i\uparrow},\hat{c}_{j\downarrow})^T$ represent the creation and annihilation operators for an electron at the $i$ site, respectively. 
By using the Bloch wave function $\ket{\epsilon_n(\bm{k})}$, this Hamiltonian $\hat{H}_0$ can be diagonalized, and the Bloch Hamiltonian $\hat{H}_{0}(\bm{k})$ in the $\bm{k}$-space is represented as
\begin{align}
\hat{H}_0(\bm{k}) 
&= t_0 (F_{\rm R}(\bm{k}) \Gamma^1 - F_{\rm Im}(\bm{k})\Gamma^{12}) + \lambda_{\rm v} \Gamma^2 \notag \\
&- \lambda_{\rm R} 
(F_1(\bm{k}) \Gamma^{4} + F_2(\bm{k}) \Gamma^{3} + F_{3}(\bm{k}) \Gamma^{24} - F_{4}(\bm{k}) \Gamma^{23})
,
\label{H0k}
\end{align}
and 
\begin{equation}
\begin{split}
F_{\rm R}(\bm{k}) &= 2\cos x \cos y + \cos 2y, \\
F_{\rm Im}(\bm{k}) &= 2\cos x \sin y - \sin 2y, \\
F_1(\bm{k}) &= \sqrt{3} \sin x \sin y, \\
F_2(\bm{k}) &= \cos x \cos y - \cos 2y, \\
F_3(\bm{k}) &= \sqrt{3} \sin x \cos y, \\
F_4(\bm{k}) &= \cos x \sin y + \sin 2y,
\end{split}
\end{equation}
with $x = \sqrt{3}k_x a_0/2$, $y = k_y a_0 /2$.
To express $H_0(\bm{k})$ in a compact form, we introduced the five Dirac matrices: $\Gamma^{1,2,3,4,5} = (\sigma_{x}\otimes s_0, \sigma_z\otimes s_0, \sigma_y \otimes s_x, \sigma_y \otimes s_y, \sigma_y \otimes s_z) $, where the Pauli matrices $\sigma_{\alpha}$ and $s_{\alpha}$ represent the sublattice index and the electron spin respectively, and $\sigma_0, s_0$ are unit matrices. 
We also introduced their ten commutators $\Gamma^{ab} = [\Gamma^{a}, \Gamma^{b}]/(2i)$.

Next, we consider a perturbation term due to the microscopic local rotation of atoms. 
Since the nearest-neighbor bond lengths change with time $t$ by the microscopic local rotations, we take these rotations as modulation of the nearest-neighbor hopping and the Rashba parameters.
We assume that the lattice deformation $\bm{u}$ modifies the nearest-neighbor hopping parameter as $t_0 \to t_0 + \delta t_{a}~(a=1,2,3)$ \cite{deformedgraphene01}. 
We define the Hamiltonian representing the modulation of the nearest-neighbor hopping term as 
\begin{equation}
\hat{H}_{\rm h} = \sum_{\braket{ij}} \delta t_{ij}(t) \hat{c}_i^\dagger \hat{c}_j.
\end{equation}
We assume that the modulation of the nearest-neighbor hopping parameter $\delta t_{a}~(a=1,2,3)$ is proportional to the change of the nearest-neighbor bond length, which is given by $\bm{u}(t)\cdot (\bm{d}_{a}/a_0)$. 
Thus, the modulation of the nearest-neighbor hopping parameter $\delta t_{a}(t)$ is given by
\begin{equation}
\delta t_a(t) = - \frac{\delta t_0}{a_0^2} \bm{u}(t)\cdot \bm{d}_a.
\end{equation}
Therefore, the Hamiltonian for the modulation of the hopping term for the wave vector $\bm{k}$ is represented as
\begin{align}
\hat{H}_{\rm h}(\bm{k},t) &= 
-\delta t (F_1(\bm{k})\Gamma^1 + F_3(\bm{k})\Gamma^{12}) \cos\Omega t \notag \\
&-\delta t(F_2(\bm{k})\Gamma^1 - F_4(\bm{k})\Gamma^{12}) \sin\Omega t,
\end{align}
with $\delta t = \delta t_0 u_{+}/a_0$. 
Similarly, we consider the modulation of the nearest-neighbor Rashba term. 
We assume that the Rashba term is modulated as
\begin{equation}
\hat{H}_{\rm R}(t) = i\lambda_{\rm R} \sum_{\braket{ij}} \hat{c}_i^\dagger \left(\bm{s}\times \frac{\bm{d}_{ij}^\prime}{|d_{ij}^\prime|}\right)_z \hat{c}_j,
\end{equation}
where $\bm{d}_{ij}^{\prime}$ is the nearest neighbor bond vector with the modulation $\bm{u}(t)$.
This can be expanded in Taylor series in powers of the displacement vector $\bm{u}$
\begin{align}
\hat{H}_{\rm R}(t) 
&= \frac{i\lambda_{\rm R}}{a_0} \sum_{\braket{ij}} \hat{c}_i^\dagger (\bm{s}\times \bm{d}_{ij})_z \hat{c}_j \notag \\
&+ \frac{-i\lambda_{\rm R}}{a_0^3}\sum_{\braket{ij}} (\bm{d}_{ij}\cdot \bm{u}(t)) \hat{c}_i^\dagger (\bm{s}\times \bm{d}_{ij})_z \hat{c}_j \notag \\
&+\frac{i\lambda_{\rm R}}{a_0}\sum_{\braket{ij}} \hat{c}_i^\dagger (\bm{s}\times \bm{u}(t))_z \hat{c}_j + O(\bm{u}^2).
\end{align}
The first term is the Rashba term without the microscopic local rotation and the second and the third terms are the modulated Rashba terms by the displacement vector $\bm{u}$. 
The modulated Rashba term $H_{\rm R}(t)$ at the wave vector $\bm{k}$ is represented as 
\begin{align}
\hat{H}_{\rm R}(\bm{k},t)
&= -\delta\lambda_{\rm R}
\Big[
F_3(\bm{k})\Gamma^{23} + F_1(\bm{k})\Gamma^{3} \notag \\
&\hspace{1cm}- (2F_{\rm Im}(\bm{k}) + 3\cos x\sin y)\Gamma^{24} \notag \\
&\hspace{1cm}+ (2F_{\rm R}(\bm{k}) + 3\cos x\cos y)\Gamma^{4}
\Big]\cos \Omega t
\notag \\
&-\delta\lambda_{\rm R}
\Big[
(2F_{\rm Im}(\bm{k}) + F_{5}(\bm{k}))\Gamma^{23} \notag \\
&\hspace{1cm}- (2F_{\rm R}(\bm{k}) + F_6(\bm{k}))\Gamma^{3} \notag \\
&\hspace{1cm}- F_3(\bm{k})\Gamma^{24} - F_1(\bm{k})\Gamma^{4}
\Big]\sin \Omega t,
\end{align}
and 
\begin{equation}
\begin{split}
F_5(\bm{k}) &= \cos x \sin y -2\sin 2y, \\
F_6(\bm{k}) &= \cos x \cos y + 2\cos 2y,
\end{split}
\end{equation}
with $\delta \lambda_{\rm R} = \frac{\lambda_{\rm R}u_+}{2a_0}$. 
Therefore, the Hamiltonian with periodic microscopic local rotation of atoms for the wave vector $\bm{k}$ is represented as 
\begin{equation}
\hat{H}(\bm{k},t) = \hat{H}_0(\bm{k}) + \hat{H}_{\rm h}(\bm{k},t) + \hat{H}_{\rm R}(\bm{k},t). \label{Hk}
\end{equation}

\section{Adiabatic series expansion}
Now we calculate spin expectation values for the time-periodic Hamiltonian. 
In general, the motions of atoms due to phonons are much slower than those of electrons because the masses of atoms are much larger than those of electrons. 
Therefore, in this section, in order to calculate the spin expectation values we employ the adiabatic series expansion in two ways: Berry-phase method in adiabatic process and iteration of unitary transformation.
The use of the adiabatic expansion is justified when the band gap is much larger than $\hbar \Omega$

\subsection{Berry-phase method in adiabatic process}
In this section, we explain the Berry-phase method to calculate dynamical response in adiabatic process proposed in Ref.~\cite{GOM}.
We will calculate an expectation values of an operator $\hat{X}(t)$.
To this end, we introduce the Hamiltonian $\hat{H}(t,\lambda)$ such that 
\begin{align}
\hat{H}(t,\lambda)|_{\lambda=0} &= \hat{H}(t) \\
\partial_{\lambda} \hat{H}(t,\lambda)|_{\lambda=0} &= \hat{X}(t). \label{EV1}
\end{align}
For example, in static systems, when the parameter $\lambda$ is an external magnetic field $B_z$, the quantity $\hat{X}$ is magnetization $M_z$. 
In particular, when the magnetic field enters the Hamiltonian as a vector potential, the quantity $\hat{X}$ is the orbital magnetization. 
On the other hand, when the magnetic field enters the Hamiltonian as a Zeeman field, $\hat{X}$ is the spin magnetization.

Next, we calculate the time-averaged expectation values of the quantity $\hat{X}(t)$, under the time-periodic Hamiltonian $\hat{H}(t)$ with time period $T$.
Let $\ket{\epsilon_{n}(t)}$ be an instantaneous eigenstate of the band $n$ at time $t$ with instantaneous eigenvalue $\epsilon_{n}(t)$, where $n$ is the band index.
We assume that time-dependence of the Hamiltonian is slow enough so that the state hardly transits to other states, which means that  the gap $\Delta_{nm}$ between bands $n$ and $m$ satisfies $\Delta_{nm} T \gg \hbar$ with time period $T$.
We introduce the Neumann equation $i\hbar \partial_t \hat{\rho}(t) = [\hat{H}(t), \hat{\rho}(t)]$ with the density matrix $\hat{\rho}(t)$.
By using the density matrix $\hat{\rho}(t)$, we define the time-average of the expectation value $\hat{X}(t)$ as
\begin{equation}
X \equiv \frac{1}{T} \int_0^T dt~{\rm tr}[\hat{\rho}(t)\hat{X}(t)]. \label{TAveEV1}
\end{equation}
We suppose that in the absence of the time evolution the density matrix is identical to $\ket{\epsilon_{n}(t)}\bra{\epsilon_{n}(t)}$ corresponding to the band $n$.
The density matrix $\hat{\rho}(t)$ can be expanded to the first order with respect to $(\Delta_{nm} T)^{-1}$ as
\begin{equation}
\hat{\rho}(t) = \ket{\epsilon_n(t)}\bra{\epsilon_{n}(t)} + \delta\hat{\rho}(t) + O((\Delta_{nm} T)^{-2}). \label{DM1}
\end{equation}
The first term is the projector onto the instantaneous eigenstate in the absence of the time evolution. 
The second term describes transitions to the excited states through the time evolution. 
The matrix element between bands $n$ and $m$ is given by 
\begin{align}
\bra{\epsilon_m(t)}\hat{\rho}(t)\ket{\epsilon_n(t)} = i\hbar \frac{\braket{\epsilon_m(t)| \partial_t \epsilon_n(t)}}{\epsilon_m(t) -\epsilon_n(t)},
\end{align}
from the Neumann equation and Eq.~(\ref{DM1}).
From Eq.~(\ref{EV1}), the matrix elements of $\hat{X}(t)$ between bands $n$ and $m$ is written as
\begin{align}
\bra{\epsilon_m(t)} \hat{X}(t) \ket{\epsilon_n(t)}
&=
\bra{\epsilon_m(t)} \partial_\lambda \hat{H}(t,\lambda)|_{\lambda=0} \ket{\epsilon_n(t)}
\notag \\
&=
(\epsilon_n(t) - \epsilon_m(t)) \braket{\epsilon_m(t) | \partial_{\lambda} \epsilon_n(t,\lambda)}|_{\lambda=0}. 
\end{align}
Here we used the Sternheimer identity $(\partial_{\lambda} \hat{H}(t,\lambda)) \ket{\epsilon_n(t,\lambda)} = (\epsilon_{n}(t,\lambda) - \hat{H}(t,\lambda))\ket{\partial_\lambda \epsilon_{n}(t,\lambda)} + \partial_{\lambda} \epsilon_{n}(t,\lambda) \ket{\epsilon_n(t,\lambda)}$. 
By combining these equations, Eq.~(\ref{TAveEV1}) is rewritten as
\begin{align}
X_{n} 
=
\frac{1}{T} \int_0^T dt  (\partial_{\lambda} \epsilon_{n}(t,\lambda) + B_{n}(t,\lambda))|_{\lambda=0},
\end{align}
where $B_{n}(t,\lambda) = i\hbar \partial_{t} \braket{\epsilon_{n}(t,\lambda) | \partial_{\lambda}\epsilon_{n}(t,\lambda)} - i\hbar \partial_{\lambda}  \braket{\epsilon_{n}(t,\lambda) | \partial_{t}\epsilon_{n}(t,\lambda)}$ is the Berry curvature of a band $n$ in the $(t,\lambda)$ space.
Moreover, in the time-periodic systems, the instantaneous eigenstates can be chosen to satisfy $\ket{\epsilon_{n}(t,\lambda)} = \ket{\epsilon_{n}(t+T,\lambda)}$.
Therefore, the time-average of the expectation value is rewritten as
\begin{align}
X_{n} &= X_{n}^{\rm inst} + X_{n}^{\rm geom},
\label{Xinst+geom}
\\
X_{n}^{\rm inst}
&\equiv
\frac{1}{T}\int_0^T dt  \bra{\epsilon_{n}(t)}\hat{X}(t) \ket{\epsilon_n(t)} \notag \\
&=
\frac{1}{T} \int_0^T dt  \partial_{\lambda}\epsilon_{n}(t,\lambda)|_{\lambda=0}, \label{Xinst}
\\
X_{n}^{\rm geom}
&\equiv
-\frac{1}{T} \partial_{\lambda} \varphi_{n}(\lambda)|_{\lambda = 0}, \label{Xgeom}
\\
\varphi_{n}(\lambda) &\equiv \int_{0}^{T} dt \braket{\epsilon_{n}(t,\lambda) | i\hbar\partial_t \epsilon_{n}(t,\lambda)},
\end{align}
where $X_{n}^{\rm inst}$ is the time-average of the expectation value of the band $n$ for the instantaneous eigenvalues, $X_{n}^{\rm geom}$ is the geometric contribution of the band $n$ due to the adiabatic time evolution, and $\varphi_{n}(\lambda)$ is the Berry phase associated with the adiabatic time evolution.

When the parameter $\lambda$ is the external magnetic field $B_z$, the time-averaged expectation values $X$ corresponds to the orbital or spin magnetization. 
$X_{n}^{\rm geom}$ in Eq.~(\ref{Xgeom}) is the geometrical magnetization, and $X_{n}^{\rm inst}$ in Eq.~(\ref{Xinst}) is the magnetization from the snapshot of the eigenstates averaged over one period. 

\subsection{Iteration of unitary transformation}
In this section, we explain the method of the adiabatic series expansion proposed by Berry \cite{irbm01, APT01}, and then we show the expectation value of each state in the second-order adiabatic approximation.

First, we consider the time-periodic Hamiltonian, $\hat{H}(t)$, and its time-dependent Schr\"odinger equation, $i\hbar \frac{d}{dt} \psi(t) = \hat{H}(t)\psi(t)$, where $\psi(t)$ is a wave function. 
This Hamiltonian is periodic in time with a period $T=2\pi/\Omega$, $\hat{H}(t+T) = \hat{H}(t)$.
Here, we introduce the rescaled time $\tau = \Omega t$, and rewrite the functions of time as $\psi(t) = \psi(\tau), \hat{H}(t) = \hat{H}(\tau), \hat{H}(\tau+2\pi) = \hat{H}(\tau)$. Hence, the time-dependent Schr\"odinger equation is rewritten as
\begin{equation}
i\hbar\Omega \frac{d}{d\tau} \psi(\tau) = \hat{H}(\tau) \psi(\tau).
\end{equation}

Next, we introduce the time-evolution operator $\hat{U}$, which is unitary. In terms of the time-evolution operator $\hat{U}(\tau)$, the wave function is written as $\psi(\tau) = \hat{U}(\tau) \psi(0)$. 
By substituting it into the Schr\"odinger equation, we get
\begin{equation}
[\hat{U}^\dagger(\tau) \hat{H}(\tau) \hat{U}(\tau) - i\hbar \Omega \hat{U}^\dagger(\tau) \partial_\tau \hat{U}(\tau)]\psi(0) = 0 \label{UT1}
\end{equation}

Here, we introduce the instantaneous eigenstate $\ket{\varepsilon_n(\tau)}$ with the instantaneous eigenvalue $\varepsilon_n(\tau)$, $\hat{H}(\tau) \ket{\varepsilon_n(\tau)} = \varepsilon_n(\tau) \ket{\varepsilon_n(\tau)}$, where $n$ is the band index.
In the following, we will calculate $\hat{U}(\tau)$ in the form $\hat{U}(\tau) = \hat{R}_0(\tau) \hat{R}_1(\tau) \hat{R}_2(\tau) \cdots$, where $\hat{R}_i(\tau)~(i=1,2,3,\dots)$ is a unitary operator.
As a zeroth-order approximation of $\hat{U}(\tau)$, we define the unitary operator $\hat{R}_0(\tau)$ as
\begin{align}
\hat{R}_0(\tau) = \sum_n \ket{\varepsilon_n(\tau)} \bra{\varepsilon_n(0)}. \label{R0}
\end{align}
Then we rewrite Eq.~(\ref{UT1}),
\begin{align}
&\hat{U}^\dagger[\hat{H}(\tau) -i\hbar \Omega \partial_\tau] \hat{U}
\notag \\
&=
\cdots \hat{R}_2^\dagger \hat{R}_1^\dagger [\hat{R}_0^\dagger \hat{H}(\tau) \hat{R}_0 -i\hbar\Omega \hat{R}_0^\dagger \partial_\tau \hat{R}_0 -i\hbar\Omega \partial_\tau]\hat{R}_1 \hat{R}_2 \cdots
\notag \\
&=
\cdots \hat{R}_2^\dagger \hat{R}_1^\dagger [\tilde{H}^{(0)}(\tau) + \Omega \tilde{V}^{(0)}(\tau) -i\hbar\Omega \partial_\tau]\hat{R}_1 \hat{R}_2 \cdots, \label{zeroth}
\end{align}
and
\begin{align}
\tilde{H}^{(0)}(\tau) 
&= 
\hat{R}_0^\dagger \hat{H}(\tau) \hat{R}_0
\notag \\
&=
\sum_n \varepsilon_{n}(\tau) \ket{\varepsilon_n(0)}\bra{\varepsilon_n(0)},
\\
\tilde{V}^{(0)}(\tau)
&=
\hat{R}_0^\dagger (-i\hbar\partial_\tau) \hat{R}_0.
\end{align}
Note that time-dependence of $\tilde{H}^{(0)}(\tau)$ is contained only in the eigenvalues $\varepsilon_n(\tau)$.

In the first-order approximation of the adiabatic series expansion, $\Omega \tilde{V}^{(0)}$ is much smaller than $\tilde{H}^{(0)}(\tau)$, and we drop the $\Omega \tilde{V}^{(0)}$ term.
Here, to the first order, we put the unitary operator $\hat{R}_2 = \hat{R}_3 = \dots = \hat{I}$ and calculate $\hat{R}_1$.
From Eq.~(\ref{zeroth}), the solution of $\tilde{H}^{(0)}(\tau)\hat{R}_1(\tau) - i\hbar\Omega \partial_\tau \hat{R}_1(\tau) = 0$ is given by
\begin{equation}
\hat{R}_1(\tau) \equiv \bar{R}_{1}(\tau) = \exp\left(\frac{1}{i\hbar\Omega} \int_0^\tau ds \tilde{H}^{(0)}(s)\right).
\end{equation}
In the first-order approximation, the state at $\tau$ is represented as
\begin{equation}
\psi(\tau) = \hat{U}(\tau)\psi(0) = \hat{R}_0(\tau) \bar{R}_1(\tau) \psi(0)
\end{equation}

Next, we consider the second-order approximation of the adiabatic series expansion.
Here, we put $\tilde{H}_1(\tau) = \tilde{H}^{(0)}(\tau) + \Omega \tilde{V}^{(0)}$, and we set the instantaneous eigenvector $\ket{\tilde{\varepsilon}_{1,n}(\tau)}$ with the instantaneous eigenvalues $\tilde{\varepsilon}_{1,n}(\tau)$, $\tilde{H}_1(\tau) \ket{\tilde{\varepsilon}_{1,n}(\tau)} = \tilde{\varepsilon}_{1,n}(\tau) \ket{\tilde{\varepsilon}_{1,n}(\tau)}$.
We perform the perturbation expansion for $\tilde{H}_1(\tau),$ and $\tilde{\varepsilon}_{1,n}(\tau)$ and $\ket{\tilde{\varepsilon}_{1,n}(\tau)}$ are represented to the first order of $\Omega$ as 
\begin{align}
\tilde{\varepsilon}_{1,n}(\tau) 
&= \varepsilon_{n}(\tau) + \Omega \bra{\varepsilon_n(0)} \tilde{V}^{(0)}(\tau) \ket{\varepsilon_n(0)}, 
\\
\ket{\tilde{\varepsilon}_{1,n}(\tau)}
&= \ket{\varepsilon_{n}(0)} 
+ \Omega \sum_{m(\neq n)} \ket{\varepsilon_{m}(0)}
\frac
{
\bra{\varepsilon_m(0)} \tilde{V}^{(0)}(\tau) \ket{\varepsilon_n(0)}
}
{
\varepsilon_n(\tau) - \varepsilon_m(\tau)
}.
\end{align}
Note that we assume that the instantaneous eigenvalues $\varepsilon_n(\tau)$ for the instantaneous eigenstates $\ket{\varepsilon_n(\tau)}$ have no degeneracy.
We define the unitary operator $\hat{R}_1(\tau)$ as
\begin{align}
\hat{R}_1(\tau) = \sum_{n} \ket{\tilde{\varepsilon}_{1,n}(\tau)}\bra{\tilde{\varepsilon}_{1,n}(0)}. \label{R1second}
\end{align}
From Eq.~(\ref{zeroth}), one has
\begin{align}
&
\hat{U}^\dagger [H(\tau) - i\hbar \Omega \partial_\tau]\hat{U}
\notag \\
&=
\cdots \hat{R}_2^\dagger [ \hat{R}_1^\dagger \tilde{H}_1(\tau) \hat{R}_1 - i\hbar \Omega \hat{R}_1^\dagger \partial_\tau \hat{R}_1 - i\hbar \partial_\tau] \hat{R}_2 \cdots
\notag \\
&=
\cdots \hat{R}_2^\dagger [ \tilde{H}^{(1)}(\tau) + \Omega \tilde{V}^{(1)}(\tau) -i\hbar \Omega\partial_\tau ] \hat{R}_2 \cdots,
\label{first}
\end{align}
and
\begin{align}
\tilde{H}^{(1)}(\tau) 
&= \hat{R}_1^\dagger \tilde{H}_1(\tau) \hat{R}_1
\notag \\
&= \sum_{n} \tilde{\varepsilon}_{1,n}(\tau) \ket{\tilde{\varepsilon}_{1,n}(0)} \bra{\tilde{\varepsilon}_{1,n}(0)},
\\
\tilde{V}^{(1)}(\tau)
&= \hat{R}_1^\dagger (-i\hbar \partial_\tau) \hat{R}_1.
\end{align}
Note that time-dependence of $\tilde{H}^{(1)}(\tau)$ is contained only in the eigenvalues $\tilde{\varepsilon}_{1,n}(\tau)$.

In the second-order approximation of the adiabatic series expansion, we assume that $\Omega \tilde{V}^{(1)}(\tau)$ is of the order $O(\Omega^2)$ and this term is sufficiently smaller than $\tilde{H}^{(1)}(\tau)$. 
Hence, we drop the $\Omega \tilde{V}^{(1)}(\tau)$ term. 
Therefore, to the second order, we put the unitary operators $\hat{R}_3 = \hat{R}_4 = \dots = \hat{I}$ and calculate $\hat{R}_2$.
From Eq.~(\ref{first}), the solution of $\tilde{H}^{(1)}(\tau) \hat{R}_2(\tau) - i\hbar \Omega \partial_\tau \hat{R}_2(\tau) = 0$ is given by
\begin{equation}
\hat{R}_2(\tau) \equiv \bar{R}_2(\tau) = \exp \left( \frac{1}{i\hbar \Omega} \int_0^\tau ds \tilde{H}^{(1)}(s) \right).
\end{equation}
In the second-order approximation, the state at time $\tau$ is represented as
\begin{equation}
\psi(\tau) = \hat{U}(\tau) \psi(0) = \hat{R}_0(\tau) \hat{R}_1(\tau) \bar{R}_2(\tau) \psi(0).
\end{equation}
Similarly, we can iterate this expansion $n$ times, and then the state at $\tau$ is given as
\begin{equation}
\psi(\tau) = \hat{R}_0(\tau) \hat{R}_1(\tau) \cdots \bar{R}_n(\tau) \psi(0).
\end{equation}

Then, the expectation value of an operator $\hat{X}$ in the second order is given by
\begin{align}
\braket{X(\tau)}_n
&= \bra{\psi_n(\tau)} \hat{X} \ket{\psi_n(\tau)}
\notag \\
&= \bra{\psi_n(0)} \bar{R}_{2}^\dagger \hat{R}_{1}^\dagger \hat{R}_0^{\dagger} \hat{X} \hat{R}_{0} \hat{R}_{1} \bar{R}_{2} \ket{\psi_n(0)}, \label{EVn}
\end{align}
where $\ket{\psi_n(0)}$ is the initial state. 
First, the unitary operator $\bar{R}_2(\tau)$ is rewritten as 
\begin{align}
\bar{R}_{2}(\tau)
=&
\exp\left(\frac{-i}{\hbar\Omega} \int_0^\tau ds \sum_n \tilde{\varepsilon}_{1,n}(s) \ket{\tilde{\varepsilon}_{1,n}(0)}\bra{\tilde{\varepsilon}_{1,n}(0)} \right) 
\notag \\
&=
\sum_n e^{\frac{-i}{\hbar\Omega}\int_0^\tau ds \tilde{\varepsilon}_{1,n}(s)} \ket{\tilde{\varepsilon}_{1,n}(0)}\bra{\tilde{\varepsilon}_{1,n}(0)}, \label{R2bar}
\end{align}
from the orthogonality relation $\braket{\tilde{\varepsilon}_{1,n}(\tau) | \tilde{\varepsilon}_{1,m}(\tau)} = \delta_{n,m}$. 
By substituting Eqs.~{(\ref{R0}), (\ref{R1second}), (\ref{R2bar})} into Eq.~(\ref{EVn}), the expectation value is rewritten as
\begin{widetext}
\begin{align}
\braket{X(\tau)}_n
&=
X_{nn}(\tau) 
+ \Omega \sum_{m (\neq n)}
\left[
\left(1 - e^{\frac{i}{\hbar\Omega} \int_0^\tau ds [\tilde{\varepsilon}_{1,n}(s)-\tilde{\varepsilon}_{1,m}(s)]} \right)
X_{nm}(\tau)
\frac{\bra{\varepsilon_{m}(0)} \tilde{V}^{(0)}(\tau) \ket{\varepsilon_{n}(0)}}
{\varepsilon_{n}(\tau) - \varepsilon_{m}(\tau)}
+
{\rm c.c.}
\right]
\notag \\
&=
X_{nn}(\tau) 
+ \hbar\Omega \sum_{m (\neq n)}
\left[
\left( 1 - e^{\frac{i}{\hbar\Omega} \int_0^\tau ds [\varepsilon_{n}(s)-\varepsilon_{m}(s) -\hbar\Omega(\gamma_{n}(s) - \gamma_{m}(s))]} \right)
X_{nm}(\tau)
\frac{\braket{\varepsilon_{m}(\tau) | (-i\partial_{\tau})\varepsilon_{n}(\tau)}}
{\varepsilon_{n}(\tau) - \varepsilon_{m}(\tau)}
+
{\rm c.c.} 
\right], \label{EVs2}
\end{align}
where $X_{nm}(\tau) = \bra{\varepsilon_n(\tau)} \hat{X} \ket{\varepsilon_m(\tau)}$, 
and $\gamma_n(\tau) = \braket{\varepsilon_n(\tau) | (-i\partial_{\tau})\varepsilon_n(\tau)}$ is Berry connection.
The first term shows the instantaneous expectation value at time $\tau$, and the second term is proportional to the inverse time period $1/T$.
The time-averaged expectation value $\bar{X}_n$ is given as 
\begin{align}
\bar{X}_n 
&= \frac{1}{2\pi} \int_0^{2\pi} d\tau \braket{X(\tau)}_n
\notag \\
&\simeq \frac{1}{2\pi} \int_0^{2\pi} d\tau 
\Bigg[
X_{nn}(\tau)
+
\hbar\Omega \sum_{m(\neq n)}
X_{nm}(\tau)
\frac{\bra{\varepsilon_m(\tau)}(-i\partial_\tau)\ket{\varepsilon_n(\tau)}}
{\varepsilon_n(\tau) - \varepsilon_m(\tau)}
+
{\rm c.c.}
\Bigg].\label{Xave}
\end{align}
\end{widetext}
Since we have assumed $\tilde{\varepsilon}_{1,n}/\hbar\Omega \gg 1$, the phase term in Eq.~(\ref{EVs2}) 
rapidly oscillates, and this phase term has been neglected after time averaging.
This time-averaged expectation value is equal to Eqs.~(\ref{Xinst}) and (\ref{Xgeom}) in the Berry-phase method in adiabatic process.

%

\section{time-averaged spin magnetization}
\begin{figure}
\includegraphics[width= 6.5cm]{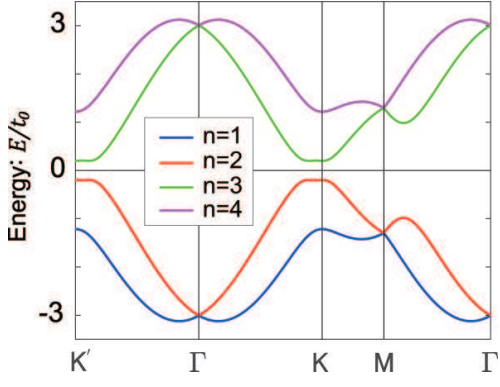}
\caption{
\label{honeyRhs}
Electronic energy band structure along the high-symmetry line in the system without phonons.
We set the parameter values to be $\lambda_{\rm v} = 0.2t_0$, and $\lambda_{\rm R} = 0.4t_0$.
}
\end{figure}

In this section, we calculate the time-averaged spin magnetization for our toy model.
The spin operator is expressed as
\begin{align}
\hat{S}_{\alpha} = \frac{\hbar}{2}\sigma_0 \otimes s_{\alpha} \equiv \frac{\hbar}{2} \hat{s}_{\alpha}
~~(\alpha = x,y,z).
\label{Sope}
\end{align}
From Eq.~(\ref{EV1}), the spin magnetization is represented as 
\begin{align}
\hat{H}_{\rm B} 
= -\hat{\bm{m}}_{\rm spin}\cdot \bm{B} 
= \frac{g\mu_{B}}{\hbar}\hat{\bm{S}}\cdot \bm{B}
= \mu_B \bm{\hat{s}} \cdot \bm{B},
\label{Mspin}
\end{align}
with g-factor $g = 2$ and the Bohr magneton $\mu_{\rm B}$.
First, we calculate the spin expectation values without phonons. 
We show the electronic energy band structure for the Hamiltonian without phonons Eq.~(\ref{H0k}) in Fig.~\ref{honeyRhs}. 
In Fig~\ref{honeyRhs}, we assume that the Fermi energy $\varepsilon_{F}$ is zero, and the temperature is zero, and then this toy model is a band insulator.
The staggered potential term breaks inversion symmetry, and the electron spin is split by the Rashba spin-orbit interaction term. 
In the following we consider the spin expectation values below the Fermi energy $\varepsilon_{F}(=0)$.

We introduce the $\Theta$ operator $\Theta = i(\sigma_0 \otimes s_y)K$, where $K$ is complex conjugate operator, and this corresponds to time-reversal operator in static systems without phonons.
The relations between the $\Theta$ operator and $\Gamma$ matrices are given by $\Theta \Gamma^{a} \Theta^{-1} = \Gamma^{a}$, $\Theta \Gamma^{ab} \Theta^{-1} = -\Gamma^{ab}$ with $a,b = 1,2,3,4,5$.
From Eq.~(\ref{H0k}), the Hamiltonian $H_0(\bm{k})$ and its eigenstate $\ket{\varepsilon_n(\bm{k})}$ with the eigenenergy $\varepsilon_{n}(\bm{k})$ is represented as $\Theta H_0(\bm{k}) \Theta^{-1} = H_0(-\bm{k})$ and $\Theta \ket{\varepsilon_n(\bm{k})} = \ket{\varepsilon_n(-\bm{k})}$.
Since the spin operator $\hat{S}_{\alpha}$ satisfies $\Theta \hat{S}_{\alpha} \Theta^{-1} = -\hat{S}_{\alpha}$, the spin expectation values $\braket{\hat{\bm{S}}(\bm{k})}_n$ is an odd function of the wavevector $\bm{k}$ in our toy model without phonons, and the total spin expectation values is zero due to cancellation between $\bm{k}$ and $-\bm{k}$.

Next, we show the spin expectation values with the microscopic local rotation of atoms.
By applying Eqs.~(\ref{Xinst+geom}), (\ref{Xinst}), and (\ref{Xgeom}) to our toy model, the time-averaged spin magnetization are given by
\begin{widetext}
\begin{align}
\bm{M}_{\rm spin}
&= \frac{1}{VT}\int_{0}^{T} dt \sum_{\bm{k}}\sum_{n}^{\rm occ} 
\Bigg[
\bra{\varepsilon_{n}(\bm{k},t)} \partial_{\bm{B}}\hat{H}_{\rm B}|_{\bm{B}=0} \ket{\varepsilon_{n}(\bm{k},t)}
\notag \\
&\hspace{2cm}
+
\sum_{m(\neq n)}
\frac
{
\bra{\varepsilon_n(\bm{k},t)} \partial_{\bm{B}} \hat{H}_{\rm B}|_{\bm{B}=0}\ket{\varepsilon_{m}(\bm{k},t)}
\braket{\varepsilon_m(\bm{k},t) | (-i\hbar\partial_t)\varepsilon_{n}(\bm{k},t)}
}
{
\varepsilon_n(\bm{k},t) - \varepsilon_m(\bm{k},t)
}
+
{\rm c.c.}
\Bigg]
\notag \\
&=
\frac{1}{2\pi} \int_0^{2\pi} d\tau \sum_{n}^{\rm occ}
[\bm{m}_{n}^{\rm inst}(\tau) + \bm{m}_{n}^{\rm geom}(\tau)],
\label{totalspinmag}
\end{align}
\end{widetext}
and
\begin{align}
\bm{m}_{n}^{\rm inst}(\tau)
&=
\frac{\mu_{\rm B}}{V}
\sum_{\bm{k}} 
\hat{\bm{s}}_{nn}(\bm{k},\tau),
\\
\bm{m}_{n}^{\rm geom}(\tau)
&=
\frac{\mu_{\rm B}}{V}
\sum_{\bm{k}} \sum_{m(\neq n)}
\Bigg[
\frac
{
\hbar \Omega \hat{\bm{s}}_{nm}(\bm{k},\tau)
A_{mn}(\bm{k},\tau)
}
{
\varepsilon_n(\bm{k},\tau) - \varepsilon_m(\bm{k},\tau)
}
+
{\rm c.c.}
\Bigg] 
\end{align}
where $\ket{\varepsilon_n(\bm{k},t)}$ is the instantaneous eigenstate with instantaneous eigenenergy $\varepsilon_n(\bm{k},t)$ for the snapshot Hamiltonian in momentum space $\hat{H}(\bm{k},t)$, $\hat{\bm{s}}_{nm}(\bm{k},\tau) = \bra{\varepsilon_n(\bm{k},\tau)} \hat{\bm{s}} \ket{\varepsilon_m(\bm{k},\tau)}$ is the instantaneous matrix element for spin between bands $n$ and $m$ at time $\tau = \Omega t$, and $A_{mn}(\bm{k},\tau) = -i\braket{\varepsilon_m(\bm{k},\tau)| \partial_{\tau} \varepsilon_n(\bm{k},\tau) }$ is the Berry connection defined by the instantaneous eigenstates between bands $n$ and $m$.
Here, the summation over $n$ represents that over the occupied bands.
$\bm{m}_{n}^{\rm inst}(\tau)$ is the instantaneous spin magnetization for the energy band $n$ at time $\tau$, and $\bm{m}_{n}^{\rm geom}(\tau)$ is the geometrical spin magnetization in the adiabatic process at time $\tau$ and this term is proportional to the phonon frequency $\Omega$.
We note that while $\hat{s}_{nm}$ and $\hat{A}_{nm}~(m\neq n)$ depend on a gauge choice, $m_{n}^{\rm geom}(\tau)$ is gauge invariant.

Here, we apply the $\Theta$ operator to our toy model with phonons.
From Eq.~(\ref{Hk}), the time-dependent Hamiltonian is rewritten as
\begin{align}
H(\bm{k},\tau) = H_0(\bm{k}) + H_c(\bm{k})\cos \tau + H_s(\bm{k})\sin \tau.
\end{align}
Therefore, the time-dependent Hamiltonian satisfies 
\begin{align}
\Theta H(\bm{k},\tau)\Theta^{-1} 
= H(-\bm{k},\tau), \label{thetaH}
\end{align}
because $\Theta H_{c(s)}(\bm{k})\Theta^{-1} = H_{c(s)}(-\bm{k})$.
Thus, within our dynamical model, the operator $\Theta$ does not flip the sign of $\tau$, and therefore, to avoid confusion we call it a $\Theta$ operation instead of time-reversal operation.
Hence the instantaneous eigenstates $\ket{\varepsilon_{n}(\bm{k},\tau)}$ and the instantaneous eigenenergy $\varepsilon_n(\bm{k},\tau)$ satisfy
\begin{align}
\Theta \ket{\varepsilon_n(\bm{k},\tau)} = \ket{\varepsilon_n(-\bm{k},\tau)},~~ \varepsilon_n(\bm{k},\tau) = \varepsilon_n(-\bm{k},\tau). \label{thetaES}
\end{align}
From these relations, the instantaneous matrix elements for spin $\hat{\bm{s}}_{nm}(\bm{k},\tau)$ and the Berry connection $A_{nm}(\bm{k},\tau)$ are given by
\begin{align}
\hat{\bm{s}}_{nm}(\bm{k},\tau) 
&= \bra{\varepsilon_n(\bm{k},\tau)} \hat{\bm{s}} \ket{\varepsilon_m(\bm{k},\tau)}
\notag \\
&= \bra{\varepsilon_n(-\bm{k},\tau)} (-\hat{\bm{s}}) \ket{\varepsilon_m(-\bm{k},\tau)}
\notag \\
&= -\hat{\bm{s}}_{nm}(-\bm{k},\tau), \label{thetas}
\end{align}
and
\begin{align}
A_{nm}(\bm{k},\tau)
&= \braket{\varepsilon_n(\bm{k},\tau) | (-i\partial_{\tau})\varepsilon_m(\bm{k},\tau)}
\notag \\
&= \braket{\varepsilon_n(-\bm{k},\tau) | (i\partial_{\tau}) \varepsilon_m(-\bm{k},\tau)}
\notag \\
&= -A_{nm}(-\bm{k},\tau). \label{thetaA}
\end{align}
By substituting these relations into Eq.~(\ref{totalspinmag}), the instantaneous spin magnetization $\bm{m}_{n}^{\rm inst}(\tau)$ becomes zero since its sum over the first Brillouin zone cancels between $\bm{k}$ and $-\bm{k}$ at time $\tau$. 
On the other hand, the geometrical spin magnetization $\bm{m}_{n}^{\rm geom}(\tau)$ can be nonzero since it is an even function of the wave vector $\bm{k}$ at time $\tau$.
Therefore, only the geometrical spin magnetization contributes to the time-averaged spin magnetization.

\begin{figure}
\includegraphics[width=8.5cm]{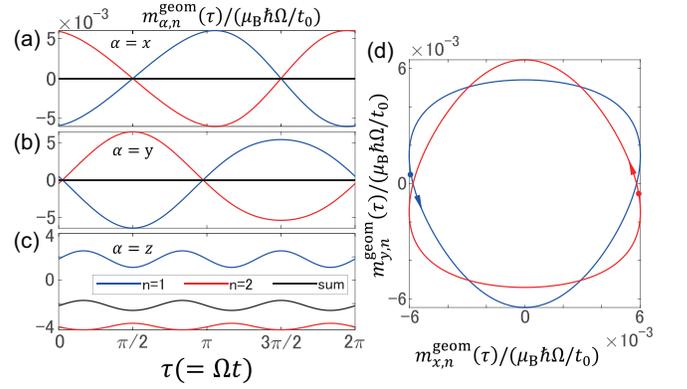}
\caption{\label{Gspin-1}
Geometrical spin magnetization for one cycle in the system with microscopic rotation versus the rescaled time $\tau$ for (a) $x$, (b) $y$, and (c) $z$ components.
(d) Evolution of spin magnetization for one cycle in the $xy$-plane. 
The small circles show the values at $\tau = 0$ and the arrows show the time-evolution direction.
We set the parameters as $\lambda_{\rm v} = 0.2t_0, \lambda_{\rm R} = 0.4t_0, \delta t = 0.1t_0$, and $\delta\lambda_{\rm R} = 0.1 \lambda_{\rm R}$.
In (a) - (d), the blue line shows the lowest band ($n=1$), and the red line shows the second lowest band ($n=2$), and the black line shows the sum of the bands below the Fermi energy.
}
\end{figure}

We calculate the geometrical spin magnetization $\bm{m}_{\alpha,n}^{\rm geom}$ for each band below the Fermi energy $\varepsilon_{F}(= 0)$, and it is nonzero as shown in Figs.~\ref{Gspin-1}(a) - (c).
In this case, the generators of the space-time symmetries are $\{C_{3z}|T/3\}$ and $\{m_\tau m_x \Theta|T/2\}$, where $\{C_{3z}|T/3\}$ is the threefold rotation around the $z$ axis, followed by $T/3$ time translation, and $\{m_{\tau}m_x\Theta|T/2\}$ is a product of  the $\Theta$ operation, the mirror reflection with respect to the $yz$-plane, and $m_\tau$ operation,  which represents inversion of the sign of time $\tau$, followed by $T/2$ time translation.
We note that the dynamical symmetry is defined using the discrete time-translation invariance Floquet systems~\cite{PhysRevB.95.195155, PhysRevLett.120.096401}.
These symmetries $\{C_{3z}|T/3\}$ and $\{m_{\tau}m_{x}\Theta|T/2\}$ are represented by unitary operators $U_1= \exp(-i\frac{\pi}{3}(\sigma_0\otimes s_z))$ and $U_2 = (\sigma_0 \otimes s_x)\Theta$, respectively. 
Here, the time-dependent Hamiltonian satisfies
\begin{align}
U_{1}\hat{H}(\bm{k},\tau)U_{1}^{-1} = \hat{H}\left(R\bm{k},\tau+ \frac{2\pi}{3} \right),
\\
U_{2} \hat{H}(k_x,k_y,\tau)U_{2}^{-1} = \hat{H}(k_x,-k_y,-\tau+\pi),
\end{align}
where $R$ is the three-fold rotational matrix around the $z$-axis, $R\bm{k} = (-k_x/2 - \sqrt{3}k_y/2, \sqrt{3}k_y/2 - k_x/2)$.
From these relations, the instantaneous matrix elements for spin $\hat{\bm{s}}_{nm}(\bm{k},\tau)$ and the Berry connection $A_{nm}(\bm{k},\tau)$ are given by
\begin{align}
\hat{\bm{s}}_{nm}(\bm{k},\tau) &= R^{-1}\hat{\bm{s}}_{nm}\left(R\bm{k},\tau+\frac{2\pi}{3}\right), \\
A_{nm}(\bm{k},\tau) &= A_{nm}\left(R\bm{k},\tau + \frac{2\pi}{3}\right),
\end{align}
and
\begin{align}
\begin{pmatrix}
\hat{s}_{x,nm}(k_x,k_y,\tau) \\
\hat{s}_{y,nm}(k_x,k_y,\tau) \\
\hat{s}_{z,nm}(k_x,k_y,\tau) 
\end{pmatrix}
&=
\begin{pmatrix}
-\hat{s}_{x,nm}(k_x,-k_y,-\tau + \pi) \\
\hat{s}_{y,nm}(k_x,-k_y,-\tau + \pi) \\
\hat{s}_{z,nm}(k_x,-k_y,-\tau + \pi) 
\end{pmatrix},
\\
A_{nm}(k_x,k_y,\tau) &= A_{nm}(k_x,-k_y,-\tau+\pi).
\end{align}
Hence, the geometrical spin magnetization satisfies
\begin{align}
\bm{m}_{n}^{\rm geom}(\tau) &= R^{-1} \bm{m}_{n}^{\rm geom}\left(\tau + \frac{2\pi}{3}\right), \\
m_{x,n}^{\rm geom}(\tau) &= -m_{x,n}^{\rm geom}(-\tau + \pi), \label{mU2x}
\\
m_{y(z),n}^{\rm geom}(\tau) &= m_{y(z),n}^{\rm geom}(-\tau+ \pi). \label{mU2yz}
\end{align}
The geometrical spin magnetization of the $z$ components oscillates with a $T/3$ period.
On the other hand, the geometrical spin magnetization in the $xy$-plane rotates, with its trajectory forming a triangle-like form for one cycle as shown in Fig~\ref{Gspin-1}(d).
From Eqs.~(\ref{mU2x}), (\ref{mU2yz}), at $\tau = \pi/2, 3\pi/2$, the geometrical spin magnetization for each band lie in the $yz$-plane.
Moreover, from the $\{C_{3z}|T/3\}$ symmetry, the geometrical spin magnetization in the $xy$-plane rotates once within the period $T$, and its time average over one period vanishes.
In addition, when the direction of the microscopic local rotation is reversed, the geometrical spin magnetizations are reversed.
Therefore, the microscopic local rotation works like an effective magnetic field, and the spin magnetization along the rotational axis is proportional to the phonon frequency.

\section{Discussion}
In this section, we discuss how the microscopic local rotation couples with electron spins.
First, when the Rashba parameter $\lambda_{\rm R}$ is zero in our toy model, the time-averaged spin magnetization is zero since the energy bands are spin degenerate and the instantaneous matrix element for spin $\hat{\bm{s}}_{nm}(\bm{k},\tau)$ becomes zero for arbitrary wave vector, time, and bands.
Therefore, within our model, the spin-orbit coupling is required for spin polarization induced by microscopic atomic rotations.

\begin{figure}
\includegraphics[width=8.5cm]{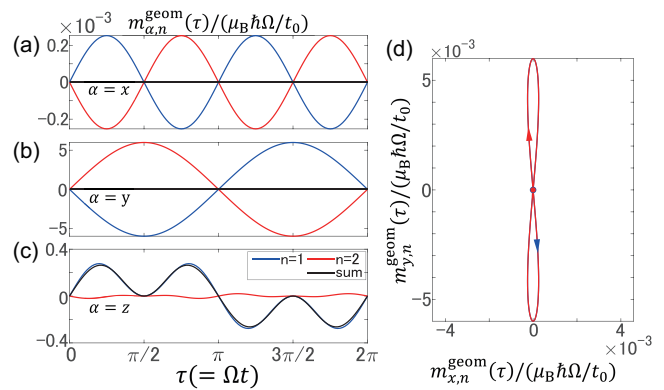}
\caption{\label{Gspin-vib1}
Geometrical spin magnetization for one cycle in the system with the simple vibration versus rescaled time $\tau$ for (a) $x$, (b) $y$, and (c) $z$ components.
(d) Evolution of spin magnetization for one cycle in the $xy$-plane. 
The small circles show the values at $\tau = 0$ and the arrows show the time-evolution direction.
We set the parameters as $\lambda_{\rm v} = 0.2t, \lambda_{\rm R} = 0.4t_0, \delta t = 0.1t_0,$ and $\delta \lambda_{\rm R} = 0.1\lambda_{\rm R}$.
In (a) - (d), the blue lines show the lowest band $(n=1)$, and the red lines show the second lowest band $(n=2)$, and the black lines show the sum of the below the Fermi energy.
}
\end{figure}

Next we compare two cases: a simple vibration of atoms and a microscopic local rotation of atoms.
For this comparison, we consider a case where the atoms A and B mutually vibrate with the displacement vector $\bm{u}$ given by
\begin{align}
\bm{u} = \bm{u}_{\rm B} - \bm{u}_{\rm A} = -u_{+}(\cos \tau, 0).
\end{align}
Then the time-dependent Hamiltonian Eq.~(\ref{Hk}) is rewritten as $H(\bm{k},\tau) = H_0(\bm{k}) + H_c(\bm{k})\cos\tau$.
We calculate the geometric spin magnetization in this system with the simple vibration of atoms, and it is shown in Fig.~\ref{Gspin-vib1}.
In Figs.~\ref{Gspin-vib1}(a), (b), and (c), we show the $x, y$ and $z$ componets of geometrical spin magnetization versus time $\tau$, respectively.
In Fig.~\ref{Gspin-vib1}(d), we show the evolution of the geometrical spin magnetization for one period in the $xy$-plane.
As is similar to the results with the microscopic local rotation of atoms, the geometrical spin magnetization can be finite at time $\tau$. 
Meanwhile the time-averaged spin magnetization vanishes.

We discuss the symmetry reason why the time-averaged spin magnetization vanishes.
In our toy model with a simple vibration of atoms, the Hamiltonian and instantaneous eigenstates satisfy not only Eqs. (\ref{thetaH}) and (\ref{thetaES}) but also the following relations, 
\begin{align}
\Theta H(\bm{k},\tau) \Theta^{-1} &= H(-\bm{k},-\tau), \label{thetamH}
\\
\Theta \ket{\varepsilon_n(\bm{k},\tau)} &= \ket{\varepsilon_n(-\bm{k},-\tau)}, \label{thetamES}
\end{align}
because $H(\bm{k},\tau) = H(\bm{k},-\tau)$.
The instantaneous matrix element for spin $\hat{\bm{s}}_{nm}(\bm{k},\tau)$ and the Berry connection $A_{nm}(\bm{k},\tau)$ satisfies
\begin{align}
\hat{\bm{s}}_{nm}(\bm{k},\tau)
&=
-\hat{\bm{s}}_{nm}(-\bm{k},-\tau), \label{vibspintheta}
\\
A_{nm}(\bm{k},\tau) 
&=
A_{nm}(-\bm{k},-\tau). \label{vibberrytheta}
\end{align}
From Eqs.~(\ref{thetas}), (\ref{thetaA}), (\ref{vibspintheta}), and (\ref{vibberrytheta}), the geometrical spin magnetization is an even function of the wave vector $\bm{k}$ and at the same time an odd function of time $\tau$, and therefore the geometrical spin magnetization has opposite signs between $\tau$ and $-\tau$.
In this case, the generators of the space-time symmetry of the system are $\{\Theta\}$ and $\{m_x| T/2\}$, which is the mirror reflection with respect to the $yz$-plane $m_x$ and the time translation by half the period $T/2$. 
This symmetry $\{m_x| T/2\}$ is represented by the unitary operator $U_3 = \sigma_0 \otimes s_x$.
From $U_3\hat{H}(k_x,k_y,\tau)U_3^{-1} = \hat{H}(-k_x,k_y,\tau+\pi)$, the instantaneous matrix element for spin and the Berry connection satisfy the following relations:
\begin{align}
\hat{s}_{x,nm}(k_x,k_y,\tau) &= \hat{s}_{x,nm}(-k_x,k_y,\tau + \pi), \label{vibspinxmirror}
\\
\hat{s}_{y(z),nm}(k_x,k_y,\tau) &= -\hat{s}_{y(z),nm}(-k_x,k_y,\tau + \pi), \label{vibspinyzmirror}
\\
A_{nm}(k_x,k_y,\tau) &= A_{nm}(-k_x,k_y,\tau+\pi). \label{vibberrymirror}
\end{align}
By combining these relations for the $\Theta$ operator (Eqs.~(\ref{vibspintheta}) and (\ref{vibberrytheta})) and symmetry operator $\{m_x|T/2\}$ (Eqs.~(\ref{vibspinxmirror}) - (\ref{vibberrymirror})), the geometrical spin magnetization $m_{\alpha,n}^{\rm geom}(\tau)$ satisfies
\begin{align}
m_{x,n}^{\rm geom}(\tau) &= -m_{x,n}^{\rm geom}(-\tau) = -m_{x,n}^{\rm geom}(-\tau + \pi),
\\
m_{y(z),n}^{\rm geom}(\tau) &= -m_{y(z),n}^{\rm geom}(-\tau) =  m_{y(z),n}^{\rm geom}(-\tau + \pi). 
\end{align}
At $\tau = 0, \pi$, the geometrical spin magnetization becomes zero, and at $\tau = \pi/2, 3\pi/2$, the $x$ component of the geometrical spin magnetization becomes zero, while its the $y$ and $z$ components are symmetric with respect to $\tau= \pi/2, 3\pi/2$.
This symmetry analysis coincides the numerical results shown in Fig.~\ref{Gspin-vib1}.
Therefore, the time-averaged spin magnetization vanishes for a simple vibration of atoms by time-reversal symmetry.


Next, we discuss an effect of the modulation of the Rashba term $\delta\lambda_{\rm R}$.
\begin{figure}
\includegraphics[width=8.5cm]{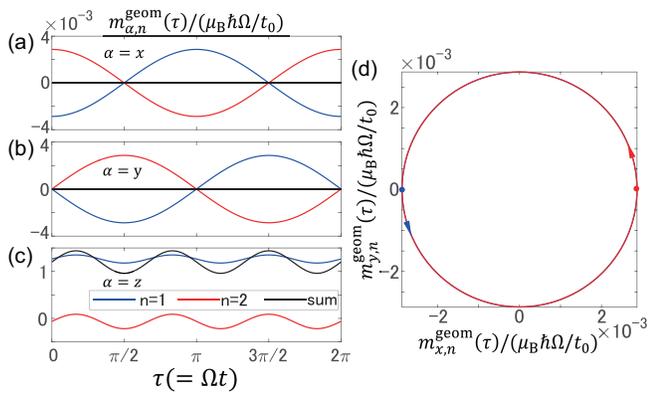}
\caption{\label{Gspin-woR}
Geometrical spin magnetization for one cycle in the system with microscopic rotation as $\delta\lambda_{\rm R} = 0$ versus the rescaled time $\tau$ for (a)$x$, (b)$y$, and (c)$z$ components.
(d) Evolution of spin magnetization for one cycle in the $xy$-plane.
The small circles show the $\tau=0$ and the arrows show the time-evolution direction.
We set the parameters as $\lambda_{\rm v} = 0.2t_0, \lambda_{\rm R} = 0.4t_0, \delta t = 0.1t_0$, and $\delta \lambda_{\rm R} = 0$.
In (a) - (d), the blue lines show the lowest band ($n=1$), and the red lines show the second lowest band ($n=2$), and the black lines show the sum of the bands below the Fermi energy.
}
\end{figure}
In our model, even without the modulation of the Rashba term $\delta \lambda_{\rm R} = 0$, the time-averaged spin magnetization along the microscopic rotational axis becomes nonzero as shown in Fig.~\ref{Gspin-woR}(c).
Figures~\ref{Gspin-woR}(a), (b), and (d) show that the geometrical spin magnetization in the $xy$-plane isotropically rotates for one cycle.
When the modulation of the Rashba term is absent, the electron spin couples with the microscopic local rotation indirectly via the Rashba spin-orbit term, and the magnitude of the geometrical spin magnetization becomes smaller than that with the modulation of the Rashba term (Fig. \ref{Gspin-1}).
Therefore, the microscopic local rotation of atoms couples with the electron spins via the Rashba spin-orbit interaction, and then the time-averaged spin magnetization along the rotational axis is generated.

\begin{figure}
\includegraphics[width=8.5cm]{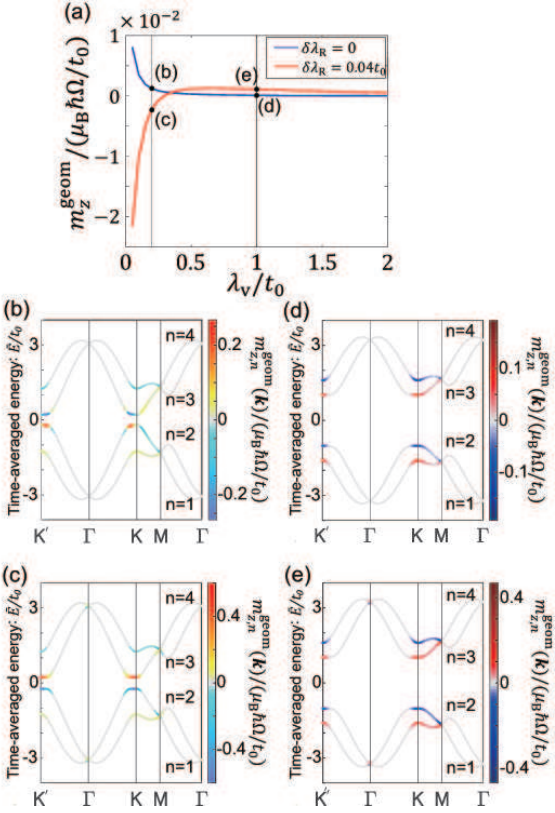}
\caption{\label{gomvsv}
Dependence on the staggered potential $\lambda_{\rm v}$ of geometrical spin magnetization.
(a) Geometrical spin magnetization with or without modulated Rashba parameter $\delta \lambda_{\rm R}$ versus staggered potential $\lambda_{\rm v}$.
The blue lines show the geometrical spin magnetization without modulated Rashba parameter $\delta \lambda_{\rm R} = 0$, and the red line show it with $\delta \lambda_{\rm R} = 0.04t$.
(b) - (e): Time-averaged energy expectation values and the geometrical spin magnetization. 
For (b), (c), (d) and (e), the colormap shows the time-averaged geometrical spin magnetization on the high-symmetry line.
We set the parameters as $\lambda_{\rm R} = 0.4t_0, \delta t = 0.1t_0$ and (b) $\lambda_{\rm v} = 0.2t_0, \delta\lambda_{\rm R} = 0$, (c) $\lambda_{\rm v} = 0.2t_0, \delta\lambda_{\rm R}= 0.04t_0$, (d) $\lambda_{\rm v} = t_0, \delta\lambda_{\rm R} = 0$, and (e) $ \lambda_{\rm v} = t_0, \delta\lambda_{\rm R} = 0.04t_0$.
}
\end{figure}

Next, we discuss how the spin magnetization depends on the gap. 
The geometrical spin magnetization can be rewritten as
\begin{align}
&\sum_{n} \bm{m}_{n}^{\rm geom}(\tau) 
\notag \\
&=
\hbar \Omega \sum_{\bm{k}}\sum_{n}^{\rm occ} \sum_{m(\neq n)}
\left[
\frac{\hat{\bm{s}}_{nm}(\bm{k},\tau)A_{mn}(\bm{k},\tau)}
{\varepsilon_n(\bm{k},\tau) - \varepsilon_m(\bm{k},\tau)}
+ {\rm c.c.}
\right]
\notag \\
&=
\hbar \Omega \sum_{\bm{k}}\sum_{n}^{\rm occ} \sum_{m}^{\rm unocc}
\left[
\frac{\hat{\bm{s}}_{nm}(\bm{k},\tau)A_{mn}(\bm{k},\tau)}
{\varepsilon_n(\bm{k},\tau) - \varepsilon_m(\bm{k},\tau)}
+ {\rm c.c.}
\right]. \label{gsm1}
\end{align}
In the second line, we divide the sum over all the bands $m$ to that over the occupied bands and that over the unoccupied bands, and the former sum vanishes.
From Eq.~(\ref{gsm1}), since the geometrical spin magnetization is proportional to the inverse of the difference of energy $[\epsilon_n(\bm{k},\tau)-\epsilon_m(\bm{k},\tau)]^{-1}$, the magnitude of the geometrical spin magnetization becomes larger when the gap is small, namely when the staggered potential $\lambda_{\rm v}$ is small, as shown in Fig.~\ref{gomvsv}(a). 
In Figs~\ref{gomvsv} (b), (c), (d) and (e), we show the the time-averaged energy expectation value $\bar{E}$ and the time-averaged geometrical spin magnetization $m_{n,z}^{\rm geom}$ on the high-symmetry line, where $\bar{E}_{n}(\bm{k}) = \frac{1}{2\pi}\int_0^{2\pi}d\tau \bra{\varepsilon_n(\bm{k},\tau)} \hat{H}(\bm{k},\tau) \ket{\varepsilon_{n}(\bm{k},\tau)}$ is the time-averaged energy expectation value and the time-averaged geometrical spin magnetization $m_{z,n}^{\rm geom}(\bm{k})$ is given from Eq.~(\ref{totalspinmag}) as
\begin{align}
&\bm{m}_{n}^{\rm geom}(\bm{k}) \notag \\
&= \frac{\mu_{\rm B}\hbar\Omega}{2\pi} \int_0^{2\pi} d\tau
\sum_{m(\neq n)}\left[
\frac{\bm{\hat{s}}_{nm}(\bm{k},\tau) A_{mn}(\bm{k},\tau)}{\varepsilon_n(\bm{k},\tau) - \varepsilon_{m}(\bm{k},\tau)} + {\rm c.c.}
\right]. \label{gsmtave}
\end{align}

\begin{figure}
\includegraphics[width=8.5cm]{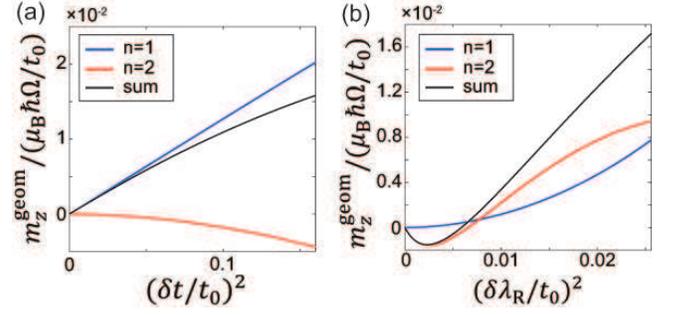}
\caption{\label{gsmvseachpara}
Dependence on the modulated hopping parameter $\delta t$ and the modulated  Rashba parameter $\delta \lambda_{\rm R}$ of the geometrical spin magnetization.
(a) Geometrical spin magnetization without the modulated Rashba parameter $\delta \lambda_{\rm R} =0$ versus the modulated hopping parameter $\delta t^2$.
(b) Geometrical spin magnetization without the modulated hopping parameter $\delta t = 0$ versus the modulated Rashba parameter $\delta \lambda_{\rm R}^2$.
We set the parameters as $\lambda_{\rm v} = 0.2t_0, \lambda_{\rm R} = 0.4t_0$.
The blue lines show the lowest band $(n=1)$, and the red lines show the second lowest band $(n=2)$, and the black lines show the sum of the bands below the Fermi energy.
}
\end{figure}

Moreover, we show the dependence on the modulated hopping and the modulated Rashba parameters of the geometrical spin magnetization in Fig~\ref{gsmvseachpara}.
From Figs.~\ref{gsmvseachpara}(a) and (b), the geometrical spin magnetization is proportional to the square of the modulated parameters $\delta t$ and $\delta \lambda_{\rm R}$ in the lowest order.
Thus the physical picture of the present effect as follows.
Our model is insulating in the static case, and the phonons dynamically excite electrons across the gap.
The excitation depends on spin via the Rashba term, and as a result, the system acquires a spin magnetization.
Figures~\ref{gomvsv} (b), (c), (d) and (e) show the distribution of the contribution to the spin magnetization in Eq.~(\ref{gsmtave}) as a color map.
When the staggered potential $\lambda_{\rm v}$ is small (Figs.~\ref{gomvsv}(b) and (c)), the gap is small, and the time-averaged geometrical spin magnetization around $K$ and $K'$ points for $n=2$ mainly contributes to the spin magnetization.
On the other hand, when the staggered potential $\lambda_{\rm v}$ becomes larger (Figs.~\ref{gomvsv}(d) and (e)), the time-averaged geometrical spin magnetization around $K$ and $K^\prime$ points cancels between $n=1$ and $n=2$ bands, and then the spin magnetization becomes small.
Moreover, since the geometrical spin magnetization is of the second order of the displacement by phonons, the spin polarization is generated by interactions between the dynamically excited states for electrons.
This clearly shows that the dynamically excited electrons across the gaps around $K$ and $K^\prime$ points generate spin polarization.
Therefore, as the band gap becomes larger, the geometrical spin magnetization becomes smaller.

\section{Conclusion}
In summary, we have theoretically show that a microscopic local rotation of atoms induces electron spins in the system with Rashba spin-orbit interaction. 
Using our toy model which is a two-dimensional tight-binding model on the honeycomb lattice with Rashba spin-orbit interaction and phonons, we calculate the time-averaged spin magnetization using the adiabatic approximation, which is valid because the phonon frequency is much smaller than the frequency scale from the electron bandwidth.
The time-averaged spin magnetization have two terms; the instantaneous spin magnetization and the geometrical spin magnetization.
The instantaneous spin magnetization is given by the instantaneous eigenstates for the snapshot Hamiltonian, and this term vanishes in the snapshot system with time-reversal symmetry.
The geometrical spin magnetization is represented in terms of the product of the Berry connection and the instantaneous matrix element for spin, and it is proportional to the phonon frequency $\Omega$.
When the system has both the Rashba spin-orbit interaction and the microscopic local rotation of atoms, the geometrical spin magnetization becomes nonzero.
The geometrical spin magnetization for each band perpendicular to the rotational axis have a finite value as a function of time, and its time average over one period is zero.
Meanwhile the time-averaged spin magnetization along the rotational axis has a finite value and it is proportional to the phonon frequency $\Omega$.
On the other hand, when the motions of atoms are vibrations, the geometrical spin magnetization for each band has a finite value as a function of time, but the time-averaged spin magnetization vanishes due to the time-reversal symmetry. 
Therefore, we have shown that the electron spins couple to the microscopic local rotation of atoms via the spin-orbit interaction, and the time-averaged spin magnetization along the rotational axis is proportional to the phonon frequency.

In the present paper, we studied phonons at the $\Gamma$ point.
We expect that phonons at general $\bm{k}$ points leads to a qualitatively similar result.
This is because the effect which we discuss is of the second order in the phonon displacement, and so even for phonons with wavevector $\bm{k}$ can give a uniform magnetization in this second order due to $\bm{k} + (-\bm{k}) = 0$.
Detailed behaviors of the spin magnetization sensitively depend on systems and phonons considered, and are left as futrue problems.

The generation of spin polarization due to rotational motion looks similar to the effect due to spin-rotation coupling discussed in Refs.~\cite{SRC01, SRC02, SRC03}.
Nonetheless there is one important difference between our work and the effects due to spin-rotation coupling.
The effect in the present requires a spin-orbit coupling, and the resulting spin polarization depends on the spin-orbit coupling, whereas the spin-rotation coupling does not depend much on spin-orbit coupling.
Thus the result is the present papaer cannot be attributed to the spin-rotation coupling in Refs.~\cite{SRC01, SRC02, SRC03}.
We also note that a relateded phenomenon is studied for an orbital magnetization in Ref.~\cite{GOM}.
It is shown that a geometric orbital magnetization appears in a periodic adiabatic process.
It does not require spin-orbit coupling.
Meanwhile, our result shows that generation of spin magnetization via rotations of phonons require spin-orbit coupling.

\begin{acknowledgments}
We thank Keiji Saito for useful discussions.
This work was partly supported by a Grant-in-Aid for Scientific Research on Innovative Area, ``Nano Spin Conversion Science'' (Grant No.~26100006), by MEXT Elements Strategy Initiative to Form Core Research Center (TIES), and also by JSPS KAKENHI Grant Numbers JP17J10342 and 18H03678.
\end{acknowledgments}





\begin{thebibliography}{35}%
\makeatletter
\providecommand \@ifxundefined [1]{%
 \@ifx{#1\undefined}
}%
\providecommand \@ifnum [1]{%
 \ifnum #1\expandafter \@firstoftwo
 \else \expandafter \@secondoftwo
 \fi
}%
\providecommand \@ifx [1]{%
 \ifx #1\expandafter \@firstoftwo
 \else \expandafter \@secondoftwo
 \fi
}%
\providecommand \natexlab [1]{#1}%
\providecommand \enquote  [1]{``#1''}%
\providecommand \bibnamefont  [1]{#1}%
\providecommand \bibfnamefont [1]{#1}%
\providecommand \citenamefont [1]{#1}%
\providecommand \href@noop [0]{\@secondoftwo}%
\providecommand \href [0]{\begingroup \@sanitize@url \@href}%
\providecommand \@href[1]{\@@startlink{#1}\@@href}%
\providecommand \@@href[1]{\endgroup#1\@@endlink}%
\providecommand \@sanitize@url [0]{\catcode `\\12\catcode `\$12\catcode
  `\&12\catcode `\#12\catcode `\^12\catcode `\_12\catcode `\%12\relax}%
\providecommand \@@startlink[1]{}%
\providecommand \@@endlink[0]{}%
\providecommand \url  [0]{\begingroup\@sanitize@url \@url }%
\providecommand \@url [1]{\endgroup\@href {#1}{\urlprefix }}%
\providecommand \urlprefix  [0]{URL }%
\providecommand \Eprint [0]{\href }%
\providecommand \doibase [0]{http://dx.doi.org/}%
\providecommand \selectlanguage [0]{\@gobble}%
\providecommand \bibinfo  [0]{\@secondoftwo}%
\providecommand \bibfield  [0]{\@secondoftwo}%
\providecommand \translation [1]{[#1]}%
\providecommand \BibitemOpen [0]{}%
\providecommand \bibitemStop [0]{}%
\providecommand \bibitemNoStop [0]{.\EOS\space}%
\providecommand \EOS [0]{\spacefactor3000\relax}%
\providecommand \BibitemShut  [1]{\csname bibitem#1\endcsname}%
\let\auto@bib@innerbib\@empty
\bibitem [{\citenamefont {Einstein}\ and\ \citenamefont
  {de~Haas}(1915)}]{Einstein}%
  \BibitemOpen
  \bibfield  {author} {\bibinfo {author} {\bibfnamefont {A.}~\bibnamefont
  {Einstein}}\ and\ \bibinfo {author} {\bibfnamefont {W.~J.}\ \bibnamefont
  {de~Haas}},\ }\href@noop {} {\bibfield {journal} {\bibinfo {journal} {Phys. Ges.}\ }\textbf {\bibinfo {volume} {17}},\ \bibinfo
  {pages} {152} (\bibinfo {year} {1915})}\BibitemShut {NoStop}%
\bibitem [{\citenamefont {Barnett}(1915)}]{Barnett}%
  \BibitemOpen
  \bibfield  {author} {\bibinfo {author} {\bibfnamefont {S.~J.}\ \bibnamefont
  {Barnett}},\ }\href {\doibase 10.1103/PhysRev.6.239} {\bibfield  {journal}
  {\bibinfo  {journal} {Phys. Rev.}\ }\textbf {\bibinfo {volume} {6}},\
  \bibinfo {pages} {239} (\bibinfo {year} {1915})}\BibitemShut {NoStop}%
\bibitem [{\citenamefont {Matsuo}\ \emph {et~al.}(2013)\citenamefont {Matsuo},
  \citenamefont {Ieda}, \citenamefont {Harii}, \citenamefont {Saitoh},\ and\
  \citenamefont {Maekawa}}]{SRC01}%
  \BibitemOpen
  \bibfield  {author} {\bibinfo {author} {\bibfnamefont {M.}~\bibnamefont
  {Matsuo}}, \bibinfo {author} {\bibfnamefont {J.}~\bibnamefont {Ieda}},
  \bibinfo {author} {\bibfnamefont {K.}~\bibnamefont {Harii}}, \bibinfo
  {author} {\bibfnamefont {E.}~\bibnamefont {Saitoh}}, \ and\ \bibinfo {author}
  {\bibfnamefont {S.}~\bibnamefont {Maekawa}},\ }\href {\doibase
  10.1103/PhysRevB.87.180402} {\bibfield  {journal} {\bibinfo  {journal} {Phys.
  Rev. B}\ }\textbf {\bibinfo {volume} {87}},\ \bibinfo {pages} {180402(R)}
  (\bibinfo {year} {2013})}\BibitemShut {NoStop}%
\bibitem [{\citenamefont {Matsuo}\ \emph
  {et~al.}(2011{\natexlab{a}})\citenamefont {Matsuo}, \citenamefont {Ieda},
  \citenamefont {Saitoh},\ and\ \citenamefont {Maekawa}}]{SRC02}%
  \BibitemOpen
  \bibfield  {author} {\bibinfo {author} {\bibfnamefont {M.}~\bibnamefont
  {Matsuo}}, \bibinfo {author} {\bibfnamefont {J.}~\bibnamefont {Ieda}},
  \bibinfo {author} {\bibfnamefont {E.}~\bibnamefont {Saitoh}}, \ and\ \bibinfo
  {author} {\bibfnamefont {S.}~\bibnamefont {Maekawa}},\ }\href {\doibase
  10.1103/PhysRevLett.106.076601} {\bibfield  {journal} {\bibinfo  {journal}
  {Phys. Rev. Lett.}\ }\textbf {\bibinfo {volume} {106}},\ \bibinfo {pages}
  {076601} (\bibinfo {year} {2011}{\natexlab{a}})}\BibitemShut {NoStop}%
\bibitem [{\citenamefont {Matsuo}\ \emph
  {et~al.}(2011{\natexlab{b}})\citenamefont {Matsuo}, \citenamefont {Ieda},
  \citenamefont {Saitoh},\ and\ \citenamefont {Maekawa}}]{SRC03}%
  \BibitemOpen
  \bibfield  {author} {\bibinfo {author} {\bibfnamefont {M.}~\bibnamefont
  {Matsuo}}, \bibinfo {author} {\bibfnamefont {J.}~\bibnamefont {Ieda}},
  \bibinfo {author} {\bibfnamefont {E.}~\bibnamefont {Saitoh}}, \ and\ \bibinfo
  {author} {\bibfnamefont {S.}~\bibnamefont {Maekawa}},\ }\href {\doibase
  10.1103/PhysRevB.84.104410} {\bibfield  {journal} {\bibinfo  {journal} {Phys.
  Rev. B}\ }\textbf {\bibinfo {volume} {84}},\ \bibinfo {pages} {104410}
  (\bibinfo {year} {2011}{\natexlab{b}})}\BibitemShut {NoStop}%
\bibitem [{\citenamefont {Kobayashi}\ \emph {et~al.}(2017)\citenamefont
  {Kobayashi}, \citenamefont {Yoshikawa}, \citenamefont {Matsuo}, \citenamefont
  {Iguchi}, \citenamefont {Maekawa}, \citenamefont {Saitoh},\ and\
  \citenamefont {Nozaki}}]{PhysRevLett.119.077202}%
  \BibitemOpen
  \bibfield  {author} {\bibinfo {author} {\bibfnamefont {D.}~\bibnamefont
  {Kobayashi}}, \bibinfo {author} {\bibfnamefont {T.}~\bibnamefont
  {Yoshikawa}}, \bibinfo {author} {\bibfnamefont {M.}~\bibnamefont {Matsuo}},
  \bibinfo {author} {\bibfnamefont {R.}~\bibnamefont {Iguchi}}, \bibinfo
  {author} {\bibfnamefont {S.}~\bibnamefont {Maekawa}}, \bibinfo {author}
  {\bibfnamefont {E.}~\bibnamefont {Saitoh}}, \ and\ \bibinfo {author}
  {\bibfnamefont {Y.}~\bibnamefont {Nozaki}},\ }\href {\doibase
  10.1103/PhysRevLett.119.077202} {\bibfield  {journal} {\bibinfo  {journal}
  {Phys. Rev. Lett.}\ }\textbf {\bibinfo {volume} {119}},\ \bibinfo {pages}
  {077202} (\bibinfo {year} {2017})}\BibitemShut {NoStop}%
\bibitem [{\citenamefont {Hamada}\ \emph {et~al.}(2015)\citenamefont {Hamada},
  \citenamefont {Yokoyama},\ and\ \citenamefont {Murakami}}]{Masato01}%
  \BibitemOpen
  \bibfield  {author} {\bibinfo {author} {\bibfnamefont {M.}~\bibnamefont
  {Hamada}}, \bibinfo {author} {\bibfnamefont {T.}~\bibnamefont {Yokoyama}}, \
  and\ \bibinfo {author} {\bibfnamefont {S.}~\bibnamefont {Murakami}},\ }\href
  {\doibase 10.1103/PhysRevB.92.060409} {\bibfield  {journal} {\bibinfo
  {journal} {Phys. Rev. B}\ }\textbf {\bibinfo {volume} {92}},\ \bibinfo
  {pages} {060409(R)} (\bibinfo {year} {2015})}\BibitemShut {NoStop}%
\bibitem [{\citenamefont {Takahashi}\ \emph {et~al.}(2016)\citenamefont
  {Takahashi}, \citenamefont {Matsuo}, \citenamefont {Ono}, \citenamefont
  {Harii}, \citenamefont {Chudo}, \citenamefont {Okayasu}, \citenamefont
  {Ieda}, \citenamefont {Takahashi}, \citenamefont {Maekawa},\ and\
  \citenamefont {Saitoh}}]{SHD01}%
  \BibitemOpen
  \bibfield  {author} {\bibinfo {author} {\bibfnamefont {R.}~\bibnamefont
  {Takahashi}}, \bibinfo {author} {\bibfnamefont {M.}~\bibnamefont {Matsuo}},
  \bibinfo {author} {\bibfnamefont {M.}~\bibnamefont {Ono}}, \bibinfo {author}
  {\bibfnamefont {K.}~\bibnamefont {Harii}}, \bibinfo {author} {\bibfnamefont
  {H.}~\bibnamefont {Chudo}}, \bibinfo {author} {\bibfnamefont
  {S.}~\bibnamefont {Okayasu}}, \bibinfo {author} {\bibfnamefont
  {J.}~\bibnamefont {Ieda}}, \bibinfo {author} {\bibfnamefont {S.}~\bibnamefont
  {Takahashi}}, \bibinfo {author} {\bibfnamefont {S.}~\bibnamefont {Maekawa}},
  \ and\ \bibinfo {author} {\bibfnamefont {E.}~\bibnamefont {Saitoh}},\ }\href
  {\doibase 10.1038/nphys3526} {\bibfield  {journal} {\bibinfo  {journal}
  {Nature Physics}\ }\textbf {\bibinfo {volume} {12}},\ \bibinfo {pages} {52}
  (\bibinfo {year} {2016})}\BibitemShut {NoStop}%
\bibitem [{\citenamefont {Zhang}\ and\ \citenamefont {Niu}(2014)}]{PAM01}%
  \BibitemOpen
  \bibfield  {author} {\bibinfo {author} {\bibfnamefont {L.}~\bibnamefont
  {Zhang}}\ and\ \bibinfo {author} {\bibfnamefont {Q.}~\bibnamefont {Niu}},\
  }\href {\doibase 10.1103/PhysRevLett.112.085503} {\bibfield  {journal}
  {\bibinfo  {journal} {Phys. Rev. Lett.}\ }\textbf {\bibinfo {volume} {112}},\
  \bibinfo {pages} {085503} (\bibinfo {year} {2014})}\BibitemShut {NoStop}%
\bibitem [{\citenamefont {Strohm}\ \emph {et~al.}(2005)\citenamefont {Strohm},
  \citenamefont {Rikken},\ and\ \citenamefont {Wyder}}]{PHEexp01}%
  \BibitemOpen
  \bibfield  {author} {\bibinfo {author} {\bibfnamefont {C.}~\bibnamefont
  {Strohm}}, \bibinfo {author} {\bibfnamefont {G.~L. J.~A.}\ \bibnamefont
  {Rikken}}, \ and\ \bibinfo {author} {\bibfnamefont {P.}~\bibnamefont
  {Wyder}},\ }\href {\doibase 10.1103/PhysRevLett.95.155901} {\bibfield
  {journal} {\bibinfo  {journal} {Phys. Rev. Lett.}\ }\textbf {\bibinfo
  {volume} {95}},\ \bibinfo {pages} {155901} (\bibinfo {year}
  {2005})}\BibitemShut {NoStop}%
\bibitem [{\citenamefont {Inyushkin}\ and\ \citenamefont
  {Taldenkov}(2007)}]{PHEexp02}%
  \BibitemOpen
  \bibfield  {author} {\bibinfo {author} {\bibfnamefont {A.~V.}\ \bibnamefont
  {Inyushkin}}\ and\ \bibinfo {author} {\bibfnamefont {A.~N.}\ \bibnamefont
  {Taldenkov}},\ }\href {\doibase 10.1134/S0021364007180075} {\bibfield
  {journal} {\bibinfo  {journal} {JETP Letters}\ }\textbf {\bibinfo {volume}
  {86}},\ \bibinfo {pages} {379} (\bibinfo {year} {2007})}\BibitemShut
  {NoStop}%
\bibitem [{\citenamefont {Sugii}\ \emph {et~al.}(2017)\citenamefont {Sugii},
  \citenamefont {Shimozawa}, \citenamefont {Watanabe}, \citenamefont {Suzuki},
  \citenamefont {Halim}, \citenamefont {Kimata}, \citenamefont {Matsumoto},
  \citenamefont {Nakatsuji},\ and\ \citenamefont {Yamashita}}]{PHEexp03}%
  \BibitemOpen
  \bibfield  {author} {\bibinfo {author} {\bibfnamefont {K.}~\bibnamefont
  {Sugii}}, \bibinfo {author} {\bibfnamefont {M.}~\bibnamefont {Shimozawa}},
  \bibinfo {author} {\bibfnamefont {D.}~\bibnamefont {Watanabe}}, \bibinfo
  {author} {\bibfnamefont {Y.}~\bibnamefont {Suzuki}}, \bibinfo {author}
  {\bibfnamefont {M.}~\bibnamefont {Halim}}, \bibinfo {author} {\bibfnamefont
  {M.}~\bibnamefont {Kimata}}, \bibinfo {author} {\bibfnamefont
  {Y.}~\bibnamefont {Matsumoto}}, \bibinfo {author} {\bibfnamefont
  {S.}~\bibnamefont {Nakatsuji}}, \ and\ \bibinfo {author} {\bibfnamefont
  {M.}~\bibnamefont {Yamashita}},\ }\href {\doibase
  10.1103/PhysRevLett.118.145902} {\bibfield  {journal} {\bibinfo  {journal}
  {Phys. Rev. Lett.}\ }\textbf {\bibinfo {volume} {118}},\ \bibinfo {pages}
  {145902} (\bibinfo {year} {2017})}\BibitemShut {NoStop}%
\bibitem [{\citenamefont {Kagan}\ and\ \citenamefont
  {Maksimov}(2008)}]{PhysRevLett.100.145902}%
  \BibitemOpen
  \bibfield  {author} {\bibinfo {author} {\bibfnamefont {Y.}~\bibnamefont
  {Kagan}}\ and\ \bibinfo {author} {\bibfnamefont {L.~A.}\ \bibnamefont
  {Maksimov}},\ }\href {\doibase 10.1103/PhysRevLett.100.145902} {\bibfield
  {journal} {\bibinfo  {journal} {Phys. Rev. Lett.}\ }\textbf {\bibinfo
  {volume} {100}},\ \bibinfo {pages} {145902} (\bibinfo {year}
  {2008})}\BibitemShut {NoStop}%
\bibitem [{\citenamefont {Zhang}\ \emph {et~al.}(2010)\citenamefont {Zhang},
  \citenamefont {Ren}, \citenamefont {Wang},\ and\ \citenamefont
  {Li}}]{PhysRevLett.105.225901}%
  \BibitemOpen
  \bibfield  {author} {\bibinfo {author} {\bibfnamefont {L.}~\bibnamefont
  {Zhang}}, \bibinfo {author} {\bibfnamefont {J.}~\bibnamefont {Ren}}, \bibinfo
  {author} {\bibfnamefont {J.-S.}\ \bibnamefont {Wang}}, \ and\ \bibinfo
  {author} {\bibfnamefont {B.}~\bibnamefont {Li}},\ }\href {\doibase
  10.1103/PhysRevLett.105.225901} {\bibfield  {journal} {\bibinfo  {journal}
  {Phys. Rev. Lett.}\ }\textbf {\bibinfo {volume} {105}},\ \bibinfo {pages}
  {225901} (\bibinfo {year} {2010})}\BibitemShut {NoStop}%
\bibitem [{\citenamefont {Garanin}\ and\ \citenamefont
  {Chudnovsky}(2015)}]{PAM02}%
  \BibitemOpen
  \bibfield  {author} {\bibinfo {author} {\bibfnamefont {D.~A.}\ \bibnamefont
  {Garanin}}\ and\ \bibinfo {author} {\bibfnamefont {E.~M.}\ \bibnamefont
  {Chudnovsky}},\ }\href {\doibase 10.1103/PhysRevB.92.024421} {\bibfield
  {journal} {\bibinfo  {journal} {Phys. Rev. B}\ }\textbf {\bibinfo {volume}
  {92}},\ \bibinfo {pages} {024421} (\bibinfo {year} {2015})}\BibitemShut
  {NoStop}%
\bibitem [{\citenamefont {Nakane}\ and\ \citenamefont
  {Kohno}(2018)}]{PhysRevB.97.174403}%
  \BibitemOpen
  \bibfield  {author} {\bibinfo {author} {\bibfnamefont {J.~J.}\ \bibnamefont
  {Nakane}}\ and\ \bibinfo {author} {\bibfnamefont {H.}~\bibnamefont {Kohno}},\
  }\href {\doibase 10.1103/PhysRevB.97.174403} {\bibfield  {journal} {\bibinfo
  {journal} {Phys. Rev. B}\ }\textbf {\bibinfo {volume} {97}},\ \bibinfo
  {pages} {174403} (\bibinfo {year} {2018})}\BibitemShut {NoStop}%
\bibitem [{\citenamefont {Streib}\ \emph {et~al.}(2018)\citenamefont {Streib},
  \citenamefont {Keshtgar},\ and\ \citenamefont
  {Bauer}}]{PhysRevLett.121.027202}%
  \BibitemOpen
  \bibfield  {author} {\bibinfo {author} {\bibfnamefont {S.}~\bibnamefont
  {Streib}}, \bibinfo {author} {\bibfnamefont {H.}~\bibnamefont {Keshtgar}}, \
  and\ \bibinfo {author} {\bibfnamefont {G.~E.~W.}\ \bibnamefont {Bauer}},\
  }\href {\doibase 10.1103/PhysRevLett.121.027202} {\bibfield  {journal}
  {\bibinfo  {journal} {Phys. Rev. Lett.}\ }\textbf {\bibinfo {volume} {121}},\
  \bibinfo {pages} {027202} (\bibinfo {year} {2018})}\BibitemShut {NoStop}%
\bibitem [{\citenamefont {Juraschek}\ and\ \citenamefont
  {Spaldin}(2019)}]{PhysRevMaterials.3.064405}%
  \BibitemOpen
  \bibfield  {author} {\bibinfo {author} {\bibfnamefont {D.~M.}\ \bibnamefont
  {Juraschek}}\ and\ \bibinfo {author} {\bibfnamefont {N.~A.}\ \bibnamefont
  {Spaldin}},\ }\href {\doibase 10.1103/PhysRevMaterials.3.064405} {\bibfield
  {journal} {\bibinfo  {journal} {Phys. Rev. Materials}\ }\textbf {\bibinfo
  {volume} {3}},\ \bibinfo {pages} {064405} (\bibinfo {year}
  {2019})}\BibitemShut {NoStop}%
\bibitem [{\citenamefont {Holanda}\ \emph {et~al.}(2018)\citenamefont
  {Holanda}, \citenamefont {Maior}, \citenamefont {Azevedo},\ and\
  \citenamefont {Rezende}}]{NatPhys.14.500}%
  \BibitemOpen
  \bibfield  {author} {\bibinfo {author} {\bibfnamefont {J.}~\bibnamefont
  {Holanda}}, \bibinfo {author} {\bibfnamefont {D.~S.}\ \bibnamefont {Maior}},
  \bibinfo {author} {\bibfnamefont {A.}~\bibnamefont {Azevedo}}, \ and\
  \bibinfo {author} {\bibfnamefont {S.~M.}\ \bibnamefont {Rezende}},\ }\href
  {\doibase doi.org/10.1038/s41567-018-0079-y} {\bibfield  {journal} {\bibinfo
  {journal} {Nature Physics}\ }\textbf {\bibinfo {volume} {14}},\ \bibinfo
  {pages} {500} (\bibinfo {year} {2018})}\BibitemShut {NoStop}%
\bibitem [{\citenamefont {Guerreiro}\ and\ \citenamefont
  {Rezende}(2015)}]{PhysRevB.92.214437}%
  \BibitemOpen
  \bibfield  {author} {\bibinfo {author} {\bibfnamefont {S.~C.}\ \bibnamefont
  {Guerreiro}}\ and\ \bibinfo {author} {\bibfnamefont {S.~M.}\ \bibnamefont
  {Rezende}},\ }\href {\doibase 10.1103/PhysRevB.92.214437} {\bibfield
  {journal} {\bibinfo  {journal} {Phys. Rev. B}\ }\textbf {\bibinfo {volume}
  {92}},\ \bibinfo {pages} {214437} (\bibinfo {year} {2015})}\BibitemShut
  {NoStop}%
\bibitem [{\citenamefont {Zhang}\ and\ \citenamefont {Niu}(2015)}]{PAM03}%
  \BibitemOpen
  \bibfield  {author} {\bibinfo {author} {\bibfnamefont {L.}~\bibnamefont
  {Zhang}}\ and\ \bibinfo {author} {\bibfnamefont {Q.}~\bibnamefont {Niu}},\
  }\href {\doibase 10.1103/PhysRevLett.115.115502} {\bibfield  {journal}
  {\bibinfo  {journal} {Phys. Rev. Lett.}\ }\textbf {\bibinfo {volume} {115}},\
  \bibinfo {pages} {115502} (\bibinfo {year} {2015})}\BibitemShut {NoStop}%
\bibitem [{\citenamefont {Zhu}\ \emph {et~al.}(2018)\citenamefont {Zhu},
  \citenamefont {Yi}, \citenamefont {Li}, \citenamefont {Xiao}, \citenamefont
  {Zhang}, \citenamefont {Yang}, \citenamefont {Kaindl}, \citenamefont {Li},
  \citenamefont {Wang},\ and\ \citenamefont {Zhang}}]{chiralphononexp01}%
  \BibitemOpen
  \bibfield  {author} {\bibinfo {author} {\bibfnamefont {H.}~\bibnamefont
  {Zhu}}, \bibinfo {author} {\bibfnamefont {J.}~\bibnamefont {Yi}}, \bibinfo
  {author} {\bibfnamefont {M.-Y.}\ \bibnamefont {Li}}, \bibinfo {author}
  {\bibfnamefont {J.}~\bibnamefont {Xiao}}, \bibinfo {author} {\bibfnamefont
  {L.}~\bibnamefont {Zhang}}, \bibinfo {author} {\bibfnamefont {C.-W.}\
  \bibnamefont {Yang}}, \bibinfo {author} {\bibfnamefont {R.~A.}\ \bibnamefont
  {Kaindl}}, \bibinfo {author} {\bibfnamefont {L.-J.}\ \bibnamefont {Li}},
  \bibinfo {author} {\bibfnamefont {Y.}~\bibnamefont {Wang}}, \ and\ \bibinfo
  {author} {\bibfnamefont {X.}~\bibnamefont {Zhang}},\ }\href {\doibase
  10.1126/science.aar2711} {\bibfield  {journal} {\bibinfo  {journal}
  {Science}\ }\textbf {\bibinfo {volume} {359}},\ \bibinfo {pages} {579}
  (\bibinfo {year} {2018})}\BibitemShut
  {NoStop}%
\bibitem [{\citenamefont {Juraschek}\ \emph {et~al.}(2017)\citenamefont
  {Juraschek}, \citenamefont {Fechner}, \citenamefont {Balatsky},\ and\
  \citenamefont {Spaldin}}]{PhysRevMaterials.1.014401}%
  \BibitemOpen
  \bibfield  {author} {\bibinfo {author} {\bibfnamefont {D.~M.}\ \bibnamefont
  {Juraschek}}, \bibinfo {author} {\bibfnamefont {M.}~\bibnamefont {Fechner}},
  \bibinfo {author} {\bibfnamefont {A.~V.}\ \bibnamefont {Balatsky}}, \ and\
  \bibinfo {author} {\bibfnamefont {N.~A.}\ \bibnamefont {Spaldin}},\ }\href
  {\doibase 10.1103/PhysRevMaterials.1.014401} {\bibfield  {journal} {\bibinfo
  {journal} {Phys. Rev. Materials}\ }\textbf {\bibinfo {volume} {1}},\ \bibinfo
  {pages} {014401} (\bibinfo {year} {2017})}\BibitemShut {NoStop}%
\bibitem [{\citenamefont {Wang}\ \emph {et~al.}(2015)\citenamefont {Wang},
  \citenamefont {Luan},\ and\ \citenamefont {Zhang}}]{Wang_2015}%
  \BibitemOpen
  \bibfield  {author} {\bibinfo {author} {\bibfnamefont {Y.-T.}\ \bibnamefont
  {Wang}}, \bibinfo {author} {\bibfnamefont {P.-G.}\ \bibnamefont {Luan}}, \
  and\ \bibinfo {author} {\bibfnamefont {S.}~\bibnamefont {Zhang}},\ }\href
  {\doibase 10.1088/1367-2630/17/7/073031} {\bibfield  {journal} {\bibinfo
  {journal} {New Journal of Physics}\ }\textbf {\bibinfo {volume} {17}},\
  \bibinfo {pages} {073031} (\bibinfo {year} {2015})}\BibitemShut {NoStop}%
\bibitem [{\citenamefont {Liu}\ \emph {et~al.}(2017)\citenamefont {Liu},
  \citenamefont {Xu}, \citenamefont {Zhang},\ and\ \citenamefont
  {Duan}}]{PhysRevB.96.064106}%
  \BibitemOpen
  \bibfield  {author} {\bibinfo {author} {\bibfnamefont {Y.}~\bibnamefont
  {Liu}}, \bibinfo {author} {\bibfnamefont {Y.}~\bibnamefont {Xu}}, \bibinfo
  {author} {\bibfnamefont {S.-C.}\ \bibnamefont {Zhang}}, \ and\ \bibinfo
  {author} {\bibfnamefont {W.}~\bibnamefont {Duan}},\ }\href {\doibase
  10.1103/PhysRevB.96.064106} {\bibfield  {journal} {\bibinfo  {journal} {Phys.
  Rev. B}\ }\textbf {\bibinfo {volume} {96}},\ \bibinfo {pages} {064106}
  (\bibinfo {year} {2017})}\BibitemShut {NoStop}%
\bibitem [{\citenamefont {Nova}\ \emph {et~al.}(2017)\citenamefont {Nova},
  \citenamefont {Cartella}, \citenamefont {Cantaluppi}, \citenamefont
  {F\"{O}rst}, \citenamefont {Bossini}, \citenamefont {Mikhaylovskiy},
  \citenamefont {Kimel}, \citenamefont {Merlin},\ and\ \citenamefont
  {Cavalleri}}]{nova2017}%
  \BibitemOpen
  \bibfield  {author} {\bibinfo {author} {\bibfnamefont {T.~F.}\ \bibnamefont
  {Nova}}, \bibinfo {author} {\bibfnamefont {A.}~\bibnamefont {Cartella}},
  \bibinfo {author} {\bibfnamefont {A.}~\bibnamefont {Cantaluppi}}, \bibinfo
  {author} {\bibfnamefont {M.}~\bibnamefont {F\"{o}rst}}, \bibinfo {author}
  {\bibfnamefont {D.}~\bibnamefont {Bossini}}, \bibinfo {author} {\bibfnamefont
  {R.~V.}\ \bibnamefont {Mikhaylovskiy}}, \bibinfo {author} {\bibfnamefont
  {A.~V.}\ \bibnamefont {Kimel}}, \bibinfo {author} {\bibfnamefont
  {R.}~\bibnamefont {Merlin}}, \ and\ \bibinfo {author} {\bibfnamefont
  {A.}~\bibnamefont {Cavalleri}},\ }\href {\doibase 10.1038/nphys3925} {\bibfield  {journal} {\bibinfo
  {journal} {Nature Physics}\  }\textbf {\bibinfo {volume} {13}},\ \bibinfo {pages} {132} (\bibinfo {year}
  {2017})}\BibitemShut {NoStop}%
\bibitem [{\citenamefont {Hamada}\ \emph {et~al.}(2018)\citenamefont {Hamada},
  \citenamefont {Minamitani}, \citenamefont {Hirayama},\ and\ \citenamefont
  {Murakami}}]{Masato02}%
  \BibitemOpen
  \bibfield  {author} {\bibinfo {author} {\bibfnamefont {M.}~\bibnamefont
  {Hamada}}, \bibinfo {author} {\bibfnamefont {E.}~\bibnamefont {Minamitani}},
  \bibinfo {author} {\bibfnamefont {M.}~\bibnamefont {Hirayama}}, \ and\
  \bibinfo {author} {\bibfnamefont {S.}~\bibnamefont {Murakami}},\ }\href
  {\doibase 10.1103/PhysRevLett.121.175301} {\bibfield  {journal} {\bibinfo
  {journal} {Phys. Rev. Lett.}\ }\textbf {\bibinfo {volume} {121}},\ \bibinfo
  {pages} {175301} (\bibinfo {year} {2018})}\BibitemShut {NoStop}%
\bibitem [{Mas()}]{Masato03}%
  \BibitemOpen
  \href@noop {} {}\bibinfo {note} {M. Hamada and S. Murakami
  (unpublished)}\BibitemShut {NoStop}%
\bibitem [{\citenamefont {Trifunovic}\ \emph {et~al.}(2019)\citenamefont
  {Trifunovic}, \citenamefont {Ono},\ and\ \citenamefont {Watanabe}}]{GOM}%
  \BibitemOpen
  \bibfield  {author} {\bibinfo {author} {\bibfnamefont {L.}~\bibnamefont
  {Trifunovic}}, \bibinfo {author} {\bibfnamefont {S.}~\bibnamefont {Ono}}, \
  and\ \bibinfo {author} {\bibfnamefont {H.}~\bibnamefont {Watanabe}},\ }\href
  {\doibase 10.1103/PhysRevB.100.054408} {\bibfield  {journal} {\bibinfo
  {journal} {Phys. Rev. B}\ }\textbf {\bibinfo {volume} {100}},\ \bibinfo
  {pages} {054408} (\bibinfo {year} {2019})}\BibitemShut {NoStop}%
\bibitem [{\citenamefont {Berry}(1987)}]{irbm01}%
  \BibitemOpen
  \bibfield  {author} {\bibinfo {author} {\bibfnamefont {M.~V.}\ \bibnamefont
  {Berry}},\ }\href {\doibase 10.1098/rspa.1987.0131} {\bibfield  {journal}
  {\bibinfo  {journal} {Proc. R. Soc. London. A}\ }\textbf {\bibinfo {volume} {414}},\
  \bibinfo {pages} {31} (\bibinfo {year} {1987})}\BibitemShut {NoStop}%
\bibitem [{\citenamefont {Rigolin}\ \emph {et~al.}(2008)\citenamefont
  {Rigolin}, \citenamefont {Ortiz},\ and\ \citenamefont {Ponce}}]{APT01}%
  \BibitemOpen
  \bibfield  {author} {\bibinfo {author} {\bibfnamefont {G.}~\bibnamefont
  {Rigolin}}, \bibinfo {author} {\bibfnamefont {G.}~\bibnamefont {Ortiz}}, \
  and\ \bibinfo {author} {\bibfnamefont {V.~H.}\ \bibnamefont {Ponce}},\ }\href
  {\doibase 10.1103/PhysRevA.78.052508} {\bibfield  {journal} {\bibinfo
  {journal} {Phys. Rev. A}\ }\textbf {\bibinfo {volume} {78}},\ \bibinfo
  {pages} {052508} (\bibinfo {year} {2008})}\BibitemShut {NoStop}%
\bibitem [{\citenamefont {Kane}\ and\ \citenamefont {Mele}(2005)}]{Z2}%
  \BibitemOpen
  \bibfield  {author} {\bibinfo {author} {\bibfnamefont {C.~L.}\ \bibnamefont
  {Kane}}\ and\ \bibinfo {author} {\bibfnamefont {E.~J.}\ \bibnamefont
  {Mele}},\ }\href {\doibase 10.1103/PhysRevLett.95.146802} {\bibfield
  {journal} {\bibinfo  {journal} {Phys. Rev. Lett.}\ }\textbf {\bibinfo
  {volume} {95}},\ \bibinfo {pages} {146802} (\bibinfo {year}
  {2005})}\BibitemShut {NoStop}%
\bibitem [{\citenamefont {Sasaki}\ and\ \citenamefont
  {Saito}(2008)}]{deformedgraphene01}%
  \BibitemOpen
  \bibfield  {author} {\bibinfo {author} {\bibfnamefont {K.-i.}\ \bibnamefont
  {Sasaki}}\ and\ \bibinfo {author} {\bibfnamefont {R.}~\bibnamefont {Saito}},\
  }\href {\doibase 10.1143/PTPS.176.253} {\bibfield  {journal} {\bibinfo
  {journal} {Prog. Theor. Phys. Suppl.}\ }\textbf {\bibinfo
  {volume} {176}},\ \bibinfo {pages} {253} (\bibinfo {year}
  {2008})}\BibitemShut {NoStop}%
\bibitem [{\citenamefont {Morimoto}\ \emph {et~al.}(2017)\citenamefont
  {Morimoto}, \citenamefont {Po},\ and\ \citenamefont
  {Vishwanath}}]{PhysRevB.95.195155}%
  \BibitemOpen
  \bibfield  {author} {\bibinfo {author} {\bibfnamefont {T.}~\bibnamefont
  {Morimoto}}, \bibinfo {author} {\bibfnamefont {H.~C.}\ \bibnamefont {Po}}, \
  and\ \bibinfo {author} {\bibfnamefont {A.}~\bibnamefont {Vishwanath}},\
  }\href {\doibase 10.1103/PhysRevB.95.195155} {\bibfield  {journal} {\bibinfo
  {journal} {Phys. Rev. B}\ }\textbf {\bibinfo {volume} {95}},\ \bibinfo
  {pages} {195155} (\bibinfo {year} {2017})}\BibitemShut {NoStop}%
\bibitem [{\citenamefont {Xu}\ and\ \citenamefont
  {Wu}(2018)}]{PhysRevLett.120.096401}%
  \BibitemOpen
  \bibfield  {author} {\bibinfo {author} {\bibfnamefont {S.}~\bibnamefont
  {Xu}}\ and\ \bibinfo {author} {\bibfnamefont {C.}~\bibnamefont {Wu}},\ }\href
  {\doibase 10.1103/PhysRevLett.120.096401} {\bibfield  {journal} {\bibinfo
  {journal} {Phys. Rev. Lett.}\ }\textbf {\bibinfo {volume} {120}},\ \bibinfo
  {pages} {096401} (\bibinfo {year} {2018})}\BibitemShut {NoStop}%
\end{thebibliography}

%







\end{document}